\documentclass[aps,prb,reprint,twocolumn,amsmath,amssymb,citeautoscript]{revtex4-2}

\usepackage{graphicx}
\usepackage{dcolumn}
\usepackage{bm}
\usepackage{amsmath}
\usepackage{amssymb}
\usepackage{xcolor}
\usepackage{hyperref}
\usepackage{natbib}
\usepackage{comment}
\usepackage{braket}
\usepackage{tikz}
\usepackage{ulem}
\usetikzlibrary{calc, arrows, positioning, decorations.pathreplacing}
\tikzset{
  spin/.style={circle,draw,inner sep=1pt,minimum size=7mm,font=\small},
  fliparrow/.style={->, thick, color=blue}
}
\DeclareGraphicsExtensions{.pdf,.png,.jpg,.jpeg,.eps}

\hypersetup{
    colorlinks=true,
    urlcolor=blue,
    citecolor=blue,
    linkcolor=blue
}

\begin{document}

\title{Tunable Floquet selection rules in a driven Ising chain}

\author{Rishi Paresh Joshi$^{1,2}$}
\email{rishiparesh.joshi@niser.ac.in}
\author{Sanchayan Banerjee$^{1,2}$}
\author{Sneha Narasimha Moorthy$^{1,2}$}
\author{Tapan Mishra$^{1,2}$}

\affiliation{$^{1}$School of Physical Sciences, National Institute of Science Education and Research, Jatni 752050, India}
\affiliation{$^{2}$Homi Bhabha National Institute, Training School Complex, Anushaktinagar, Mumbai 400094, India}

\date{\today}

\begin{abstract}
We study a periodically driven spin-$1/2$ Ising chain with a nearest-neighbour coupling and longitudinal field while a weak transverse field induces single-spin flips. Through Floquet perturbation theory (FPT), we obtain signatures of Hilbert space fragmentation (HSF) and an unconventional form of dynamical localisation which we call the Floquet freezing. Our analysis suggests that these observations emerge due to a single Floquet selection rule that dictates the prethermal dynamics. For a special value of the field-to-interaction strength ratio together with commensurate drive periods, this rule permits only a constrained subset of bulk spin flips, leading to prethermal HSF in the full spin-$1/2$ Hilbert space. Under open boundary conditions, the same rule suppresses boundary spin flips up to higher order in perturbation and produces long-lived prethermal edge memory, which is neither topological in origin nor is a strong zero mode. Furthermore, under periodic boundary conditions, the largest surviving fragment is exactly the PXP sector at leading order and therefore exhibits Floquet-inherited scar phenomenology in the prethermal window. At higher commensurate ratios of field strength to interaction strength, all first-order single-spin-flip channels are suppressed and the system enters a regime of Floquet freezing. Hence, our study leverages the selection rules obtained through Floquet perturbation theory to obtain exotic prethermal phenomena at different parameter regimes.

\end{abstract}

\maketitle

\section{Introduction}
\label{sec:intro}
Periodically driven many-body systems provide a controlled route to engineer effective stroboscopic dynamics. Generic interacting drives, however, absorb energy and can approach featureless long-time states. Avoiding such heating is therefore a central constraint in Floquet engineering~\cite{DAlessioPolkovnikov2013AnnPhys,DAlessioKafriPolkovnikovRigol2016AdvPhys}. One broadly useful route in this context is high-frequency driving. In this regime, heating can be parametrically slow, and an effective Floquet Hamiltonian can govern a long prethermal window~\cite{Abanin2015PRL,Abanin2017PRB,Kuwahara2016AnnPhys,Mori2016PRL,HoMoriAbaninDallaTorre2023AnnPhys}. Such windows have been used to realize dynamical stabilization and controllable effective Hamiltonians~\cite{Eckardt2017RMP,OkaKitamura2019ARCMP}. They have also been used to generate driven topological responses and time-translation symmetry breaking~\cite{ElseNayak2016PRB,VonKeyserlingkSondhi2016PRB1,PotterMorimotoVishwanath2016PRX,ElseBauerNayak2016PRLTimeCrystals,HarperRoyRudnerSondhi2020ARCMP,ElseBauerNayak2017PRX}. These prethermal Floquet windows not only delay heating, rather they also create a natural setting in which many exotic nonergodic many-body dynamics can emerge.

A distinct mechanism for nonergodic dynamics is Hilbert-space fragmentation (HSF)~\cite{Sala2020PRX,Khemani2020PRBShattering,MoudgalyaMotrunich2022PRX,MoudgalyaBernevigRegnault2022RPP,Mukherjee2021PRBMinimalHSF,AdityaDharSen2024PRB}. Existing clean driven realizations of HSF already span several settings. One setting is provided by periodically driven fermionic chains, where prethermal HSF has been demonstrated~\cite{Ghosh2023PRL,Ghosh2024PRB}. A second setting arises in Stark lattices, where the drive can again generate fragmented dynamics~\cite{Zhang2024StarkHSF}. Related physics also appears in homogeneous central-spin settings, where fragmentation supports nonthermal subspace dynamics and time-crystalline responses~\cite{Kumar2025CentralSpinHSF,Tang2025DTCviaHSF}. Beyond such model-specific settings, recent work has identified interference-based routes to drive-induced constraints, showing that prethermal fragmentation can emerge without imposing constraints by hand~\cite{GhoshPaulSengupta2025Spin}. More broadly, these developments connect to emergent symmetries and structures in prethermal Floquet dynamics~\cite{BanerjeeSengupta2025JPCM}.

Prethermal HSF systems often contain dynamically inactive sectors or frozen configurations, suggesting a natural connection to the broader understanding of dynamical freezing~\cite{Das2010PRBFreezing,Bhattacharyya2012PRB} or dynamical localisation~\cite{AgarwalaSen2017PRB,MondalPekkerSengupta2012EPL}. In this context, dynamical freezing refers to a dynamical arrest of particles arising from an approximate suppression of motion and preserves memory of local observables~\cite{Das2010PRBFreezing,Bhattacharyya2012PRB,Hegde2014PRB,Haldar2021PRX,Haldar2024ThermoLimit}. Closely related studies have addressed dynamical localization which arises from an interference effect that suppresses the entire dynamics in periodically driven models~\cite{AgarwalaSen2017PRB,MondalPekkerSengupta2012EPL,NagRoyDuttaSen2014PRB,NandySenguptaSen2018JPhysA,MakkiBandyopadhyayMaityDutta2022PRB,AdityaSen2023SciPostCore}. More recently, the eventual breakdown of freezing near commensurate driving has been analyzed through nonperturbative effects~\cite{MukherjeeGuoChowdhury2026PRX,LongCrowleyKhemaniChandran2023PRL}. These observations motivate a central question for the present work: whether the constrained dynamics characteristic of HSF and suppression of the motion associated with dynamical freezing or dynamical localization can, in driven systems, arise from a common underlying mechanism. This in turn raises a more specific question: whether long-lived boundary memory can emerge directly from the same driving-induced local constraint that governs the bulk dynamics, without invoking topological edge-mode mechanisms~\cite{ElseNayak2016PRB,HarperRoyRudnerSondhi2020ARCMP,ParkerVasseurScaffidi2019PRL} or strong-zero-mode physics~\cite{ElseFendleyKempNayak2017PRX,KempYaoLaumann2020PRL,YatesMitra2022CommunPhys}. More broadly, can such constrained driven dynamics be extended beyond one-dimensional chains to higher dimensions and different lattice geometries?

Motivated by the above questions, in this work, we study a periodically driven spin-$1/2$ Ising chain by considering a Floquet driving protocol where Ising interaction and the longitudinal field reverse sign halfway through each drive period. In addition to this, we consider a weak transverse field which flips single-spins. Using interaction picture Floquet perturbation theory (FPT), we derive the first order Floquet Hamiltonian and the constraints in this model~\cite{SenSenSengupta2021JPCM,GhoshPaulSengupta2025Spin,ShevchenkoAshhabNori2010PhysRep}. Our analysis reveals that at commensurate driving periods, only certain spin-flips survive, while others are suppressed. For specific values of the field to interaction strength ratio, we obtain prethermal HSF regime in the full spin-$1/2$ Hilbert space. Further, we find that under periodic boundary conditions, the largest fragment reduces to the PXP sector at first order in perturbation. Thus, the driven system  showcases the PXP-type scar phenomenology within the prethermal window, rather than through explicit microscopic PXP construction~\cite{Bernien2017Nature,Turner2018NatPhys,Mukherjee2020PRBFloquetScars,Giudici2024PRL,Ghosh2026arXiv,HudomalDesaulesMukherjeeSuHalimehPapic2022PRB,MizutaTakasanKawakami2020PRResearch,SugiuraKuwaharaSaito2021PRResearch,PaiPretko2019PRL}. Moreover at higher commensurate ratios of field to interaction strength, the first-order Floquet Hamiltonian vanishes resulting in the system entering a regime where the entire prethermal dynamics is frozen, which we term as Floquet freezing. However, under open boundary condition we obtain additional structured constraints which gives rise to long-lived boundary memory. Finally, we extend the FPT construction to higher dimensions and different lattice geometries.

\section{Model}
\label{sec:model}
We study a periodically driven Ising chain with a symmetric square-pulse sign-flip protocol of its diagonal part. The Hamiltonian contains a diagonal time-dependent Ising interaction and longitudinal part $H_0(t)$, and a weak (perturbative) static
transverse spin-flip term $H_1$, 
\begin{equation}
    \begin{aligned}
H(t)&=H_0(t)+H_1,\\ \nonumber
H_0(t)&=\Lambda(t)\bigg(-J_0\sum_{i=1}^{L_b}\sigma_i^z\sigma_{i+1}^z
-h_0\sum_{i=1}^{L}\sigma_i^z\bigg),\\ \nonumber
H_1&=-g\sum_{i=1}^{L}\sigma_i^x,
\label{eq:model_H}
\end{aligned}
\end{equation}
which represents a spin-$1/2$ chain of length $L$, where
$\{\sigma_i^x,\sigma_i^y,\sigma_i^z\}$ are the Pauli operators on site
$i$. The parameter $g$ is weak, with $|g|\ll \{|J_0|,|h_0|\}$. We work with the $\sigma^z$-basis as the computational basis. Here
$L_b=L-1$ for open boundary conditions (OBC) and $L_b=L$ for periodic
boundary conditions (PBC), with $\sigma_{L+1}^z\equiv \sigma_1^z$ in the
periodic case. We use units $\hbar=1$ throughout.

The Ising coupling and longitudinal field follow the same symmetric
square-wave sign-flip protocol:
\begin{equation}
\Lambda(t)=
\begin{cases}
+1, & 0\le t<T/2,\\
-1, & T/2\le t<T,
\end{cases}
\label{eq:model_drive}
\end{equation}
The driven part is diagonal in the computational ($\sigma^z$) basis and
reverses sign exactly halfway through the cycle of time period, $T$. This structure is most
conveniently expressed by introducing the diagonal operator
\begin{equation}
\mathcal{O}:=
J_0\sum_{i=1}^{L_b}\sigma_i^z\sigma_{i+1}^z
+h_0\sum_{i=1}^{L}\sigma_i^z.
\label{eq:model_O}
\end{equation}
In the $\sigma^z$ basis, $\mathcal{O}$ measures the magnitude of the diagonal energy cost.
A spin flip induced by $H_1$ connects two basis states with different
eigenvalues of $\mathcal{O}$. That eigenvalue change controls the
corresponding matrix element of the effective Floquet Hamiltonian. In
terms of $\mathcal{O}$, the driven piece is simply $H_0(t)=-\mathcal{O}$
in the first half-cycle and $H_0(t)=+\mathcal{O}$ in the second
half-cycle.

Because $\mathcal{O}$ is time independent, $H_0(t)$ commutes with itself
at all times. The unperturbed propagator can therefore be written
exactly as
\begin{equation}
U_0(t,0)=\exp\!\bigl[i\,f(t)\,\mathcal{O}\bigr],~
f(t)=
\begin{cases}
t, & 0\le t\le T/2,\\
T-t, & T/2<t\le T,
\end{cases}
\label{eq:model_U0}
\end{equation}
Here $f(t)$ is a triangular micromotion function. It increases linearly
during the first half of the drive and decreases linearly back to zero in
the second half. As a result, $f(T)=0$, so the unperturbed micromotion
closes exactly after one period and $U_0(T,0)=\mathbb{I}$.

This exact closure simplifies the interaction picture FPT expansion. Since the unperturbed evolution returns to the identity after one period, the full
Floquet operator is generated entirely by the interaction-picture
perturbation. Thus, the time-evolution operator for the complete Hamiltonian  for one period is $U(T,0)=U_{0}(T,0)U_I(T,0):=U_F$ with $U_F=e^{-iH_FT}$, the Floquet generator defining the Floquet Hamiltonian
$H_F$. The perturbative expansion is controlled by the small ratio
$g/\{J_0,h_0\}$. At leading order, the Floquet generator is the time
average of the interaction-picture perturbation over one period:
\begin{equation}
    \begin{aligned}
V_I(t)&:=U_0^\dagger(t,0)\,H_1\,U_0(t,0),\\ \nonumber
U_F&\equiv U_{I}(T,0)
=\mathcal{T}\exp\!\left[-i\int_0^T V_I(t)\,dt\right],\\ \nonumber
H_F^{(1)}&=\frac{1}{T}\int_0^T V_I(t)\,dt,
\label{eq:model_Floquet}
\end{aligned}
\end{equation}

In this paper, we obtain the first-order Floquet Hamiltonian in closed
form for both OBC and PBC in the computational basis. Its matrix elements
acquire an interference factor that acts as a selection rule for
single-spin-flip processes. We derive that first-order generator
explicitly in the next section.

\section{Results}

The results are organized around the first-order Floquet generator, since it already contains the local selection rule that controls the leading prethermal dynamics. We first derive this generator in closed form and identify the interference factor that dresses each single-spin-flip process. We then use that result to explain the three regimes that structure the rest of the paper: constrained bulk dynamics at the fragmentation point, boundary suppression under OBC, and complete first-order freezing at higher commensurate fields. We end our analysis with generalising the FPT construction for a periodically driven spin-$1/2$ Ising chain with driving protocol where Ising interaction and the longitudinal field reverse sign halfway through each drive period, in the presence of a static transverse field which flips single spins, to higher dimensions and different lattice geometries. 

\subsection{Effective Floquet Hamiltonian and the selection rules for spin-flip processes}
\label{sec:HF}

To understand the mechanism underlying the constrained dynamics, we first evaluate the first-order Floquet Hamiltonian matrix elements which correspond to a single-spin-flip process in the computational basis. For the present sign-flip protocol, this quantity can be obtained analytically in closed form.
 Since
\begin{equation}
H_1=-g\sum_{i=1}^{L}\sigma_i^x,
\end{equation}
the perturbation connects only pairs of basis states that differ by one spin flip, hence the first-order Floquet Hamiltonian has support only on these single-site spin flip states. For each such pair, the matrix element is multiplied by a finite-time interference factor set by the change in the eigenvalue of the diagonal operator $\mathcal{O}$ between the two connected states, equivalently by the local quantity $\Delta P_i(s)$. The derivation is straightforward: we evaluate the first-order Floquet Hamiltonian in the $\mathcal{O}$ eigenbasis and read off the resulting filter for each channel. Related interference-based selection rules are familiar in Floquet settings~\cite{Bukov2015AdvPhys,SenSenSengupta2021JPCM,Ghosh2023PRL,Ghosh2024PRB,GhoshPaulSengupta2025Spin,Ghosh2026arXiv}, but in the present interacting spin chain the result takes a simple local form. This single rule already anticipates the main results that follow: constrained bulk dynamics, edge exclusion under OBC, and the freezing regime.
\begin{figure}
    \centering
    \includegraphics[width=\linewidth]{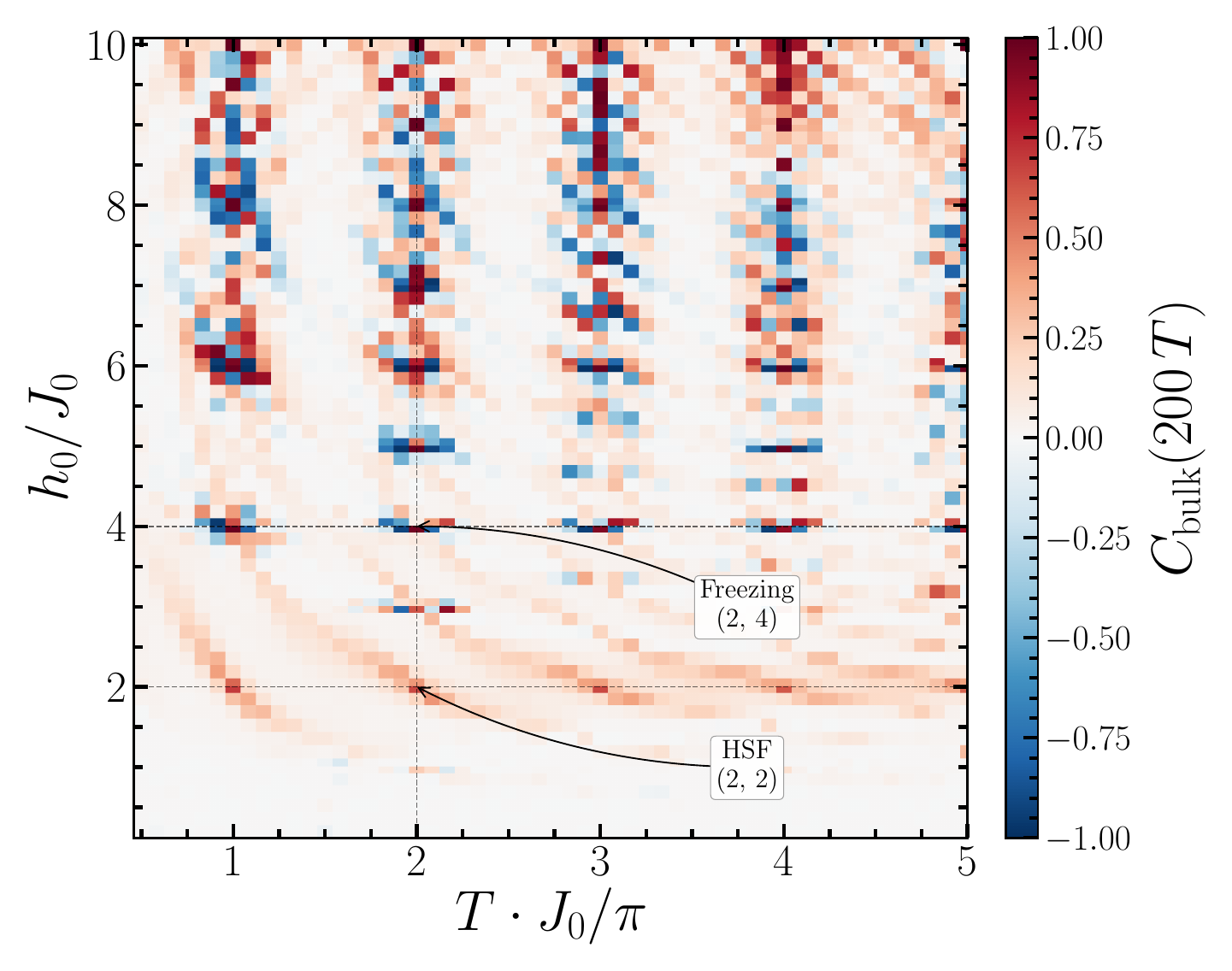}
    \caption{Late-time prethermal stroboscopic bulk-site autocorrelation, $C_{\mathrm{bulk}}(200T)$, in the dimensionless parameter
plane $(T J_0/\pi,\; h_0/J_0)$ for the periodically driven Ising chain.
The neighbourhood of $(T J_0/\pi,h_0/J_0)=(2,2)$ identifies the
prethermal HSF regime, where only the $\Delta P=0$ bulk channels
survive at first order. Thin dashed lines mark the analytically
predicted selection-rule conditions $h_0=2J_0$, $h_0=4J_0$, and
$T=2\pi/J_0$, and the arrows indicate the parameter cuts used in the
time-trace figures. Red and blue denote positive and negative late-time
memory, respectively, while white corresponds to
$C_{\mathrm{bulk}}\approx 0$. The data are for a chain of length
$L=14$, with $g=1$ and $J_0=10$, averaged over $10$ random
computational-basis initial states. The plot is shown on a $69\times 60$
grid without interpolation.}
\label{fig:phase_map}
\end{figure}
The overall structure of the prethermal regimes is already visible in Fig.~\ref{fig:phase_map}. In particular, the neighbourhood of $(T J_0/\pi,h_0/J_0)=(2,2)$ corresponds to the HSF regime, while the neighbourhood of $(2,4)$ corresponds to complete first-order freezing. The purpose of this section is to identify the local channel rule underlying this map.

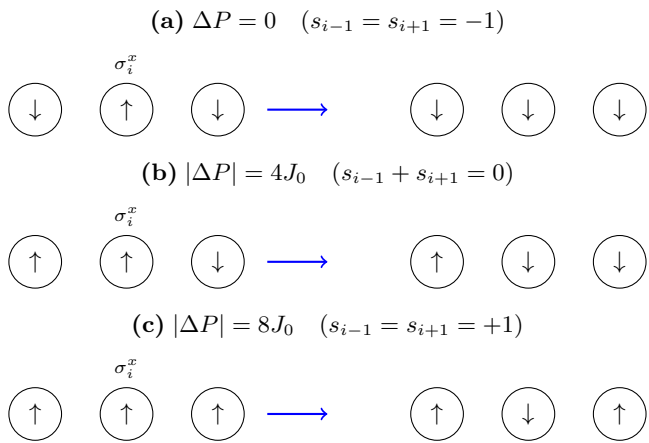
\begin{figure}[t]
\centering

\textbf{(a)}\;$\Delta P=0$\quad
($s_{i-1}=s_{i+1}=-1$)\\[4pt]
\begin{tikzpicture}
  \node[spin] (L) {$\downarrow$};
  \node[spin, right=5mm of L, label=above:{\scriptsize $\sigma_i^x$}] (M) {$\uparrow$};
  \node[spin, right=5mm of M] (R) {$\downarrow$};
  \draw[fliparrow] ($(R.east)+(3mm,0)$) -- ++(8mm,0);
  \node[spin, right=22mm of R] (L2) {$\downarrow$};
  \node[spin, right=5mm of L2] (M2) {$\downarrow$};
  \node[spin, right=5mm of M2] (R2) {$\downarrow$};
\end{tikzpicture}

\vspace{6pt}

\textbf{(b)}\;$|\Delta P|=4J_0$\quad
($s_{i-1}+s_{i+1}=0$)\\[4pt]
\begin{tikzpicture}
  \node[spin] (L) {$\uparrow$};
  \node[spin, right=5mm of L, label=above:{\scriptsize $\sigma_i^x$}] (M) {$\uparrow$};
  \node[spin, right=5mm of M] (R) {$\downarrow$};
  \draw[fliparrow] ($(R.east)+(3mm,0)$) -- ++(8mm,0);
  \node[spin, right=22mm of R] (L2) {$\uparrow$};
  \node[spin, right=5mm of L2] (M2) {$\downarrow$};
  \node[spin, right=5mm of M2] (R2) {$\downarrow$};
\end{tikzpicture}

\vspace{6pt}

\textbf{(c)}\;$|\Delta P|=8J_0$\quad
($s_{i-1}=s_{i+1}=+1$)\\[4pt]
\begin{tikzpicture}
  \node[spin] (L) {$\uparrow$};
  \node[spin, right=5mm of L, label=above:{\scriptsize $\sigma_i^x$}] (M) {$\uparrow$};
  \node[spin, right=5mm of M] (R) {$\uparrow$};
  \draw[fliparrow] ($(R.east)+(3mm,0)$) -- ++(8mm,0);
  \node[spin, right=22mm of R] (L2) {$\uparrow$};
  \node[spin, right=5mm of L2] (M2) {$\downarrow$};
  \node[spin, right=5mm of M2] (R2) {$\uparrow$};
\end{tikzpicture}

\caption{Bulk single-spin-flip channels for a middle site $i$ with two
neighbors, at $h_0=2J_0$.
(a) Both neighbors down: $\Delta P_i=0$, so this channel survives the
sinc filter for all $T$.
(b) Mixed neighbors: $|\Delta P_i|=4J_0$, so this channel is suppressed
at $T=\pi/J_0$.
(c) Both neighbors up: $|\Delta P_i|=8J_0$, so this channel is
suppressed at $T=\pi/(2J_0)$.
At the canonical period $T=2\pi/J_0$, channels (b) and (c) lie on sinc
zeros, leaving only the $\Delta P=0$ channel in panel (a).}
\label{fig:bulk_flips}
\end{figure}

For $\mathcal{O}\ket{m}=P_m\ket{m}$, the first-order Floquet matrix element between basis states $\ket{m}$ and $\ket{n}$ is
\begin{eqnarray}
\bigl(H_F^{(1)}\bigr)_{nm}
&=
(H_1)_{nm}\,
\operatorname{sinc}\!\left(\frac{P_{nm}T}{4}\right)
e^{-iP_{nm}T/4},\\ \nonumber
&P_{nm}:=P_n-P_m .
\label{eq:HF1_sinc_main}
\end{eqnarray}
Here $H_F^{(1)}$ is the first-order Floquet Hamiltonian, $(H_1)_{nm}$ is the bare single-site spin-flip matrix element of the transverse perturbation, $P_{nm}$ is the difference of $\mathcal{O}$-eigenvalues between the two basis states, and $T$ is the drive period. The $\operatorname{sinc}$ factor is the essential object. It arises because the perturbation accumulates a phase during the first half-cycle and then partially unwinds it during the second half-cycle; the residual amplitude after one full period is therefore a finite-time (prethermal) destructive-interference factor. Equation~\eqref{eq:HF1_sinc_main} immediately shows what the drive does. If $P_{nm}=0$, the channel survives with its full first-order weight. If $P_{nm}T/4=\pi k$ with nonzero integer $k$, the channel is deleted. The drive therefore does not merely renormalize matrix elements; it removes selected channels altogether \cite{GhoshPaulSengupta2025Spin}.
To translate this into local spin language (Eq.~\eqref{eq:HF1_sinc_main}), consider a computational-basis configuration $\ket{s}=\ket{s_1,\dots,s_L}$ with $s_i=\pm 1$, and let $\ket{s^{(i)}}=\sigma_i^x\ket{s}$ denote the state obtained by flipping spin $i$. The $P_{nm}$ joining the above states is:
\begin{equation}
\Delta P_i(s):=P\!\bigl(s^{(i)}\bigr)-P(s),
\label{eq:DeltaP_main}
\end{equation}
namely the change in the diagonal $\mathcal{O}$-eigenvalue caused by that single-spin flip. For a bulk site $2\le i\le L-1$, one finds $\Delta P_i(s)=-2s_i\!\left[J_0(s_{i-1}+s_{i+1})+h_0\right]$, whereas for the left edge of an open chain, $\Delta P_1(s)=-2s_1(J_0s_2+h_0)$. The local channel content at $h_0=2J_0$ is shown schematically in Figs.~\ref{fig:bulk_flips} and \ref{fig:edge_flips}. In the bulk, the allowed magnitudes are $|\Delta P_i|=0$, $4J_0$, and $8J_0$. At the edge, the allowed magnitudes become $|\Delta P_1|=2J_0$ and $6J_0$. This coordination dependence is the microscopic reason open and periodic chains differ already at first order.

The driving time period regime now follows immediately (see Tab.~\ref{tab:sinc_OBC}). At the primary resonance
$h_0=2J_0$ and $T=2\pi/J_0$, only the bulk channels with $\Delta P_i=0$ survive, while all nonzero bulk and edge single-spin-flip channels sit on zeros of the sinc filter. The surviving first-order Hamiltonian is therefore the PXP-type constrained generator \begin{equation}
    H_F^{(1)}=-g\sum_{i=2}^{L-1}\pi_{i-1}\sigma_i^x\pi_{i+1},
    \label{eq:Heff_PXP_OBC}
\end{equation} 
for OBC, with the periodic extension in PBC with $i\in\{1, \cdots,L\}$. We will generalise the absence of edge flip terms to any regular graph where the ratio becomes $z$, the bulk coordination number. By contrast, at $h_0=2nJ_0$ with integer $n>1,~n\in \mathbb{Z}$ and the same period $T=2\pi/J_0$, every first-order single-spin-flip channel is pushed onto a nonzero sinc zero, and one obtains $H_F^{(1)}=0$. HSF and freezing are therefore not independent mechanisms in this model. They are two distinct parameter regimes of the same first-order interference filter.

Following this result the surviving channels are identified and the Hilbert-space structure, the boundary hierarchy under OBC, and the numerical diagnostics follow from the same local selection rule.

\begin{figure}[t]
\centering

\textbf{(a)}\;$|\Delta P|=2J_0$\quad
($s_2=-1$, neighbor down)\\[4pt]
\begin{tikzpicture}
  \node[spin, label=above:{\scriptsize $\sigma_1^x$}] (L) {$\uparrow$};
  \node[spin, right=5mm of L] (M) {$\downarrow$};
  \node[spin, right=5mm of M] (R) {$\downarrow$};
  \draw[fliparrow] ($(R.east)+(3mm,0)$) -- ++(8mm,0);
  \node[spin, right=22mm of R] (L2) {$\downarrow$};
  \node[spin, right=5mm of L2] (M2) {$\downarrow$};
  \node[spin, right=5mm of M2] (R2) {$\downarrow$};
\end{tikzpicture}

\vspace{6pt}

\textbf{(b)}\;$|\Delta P|=6J_0$\quad
($s_2=+1$, neighbor up)\\[4pt]
\begin{tikzpicture}
  \node[spin, label=above:{\scriptsize $\sigma_1^x$}] (L) {$\uparrow$};
  \node[spin, right=5mm of L] (M) {$\uparrow$};
  \node[spin, right=5mm of M] (R) {$\downarrow$};
  \draw[fliparrow] ($(R.east)+(3mm,0)$) -- ++(8mm,0);
  \node[spin, right=22mm of R] (L2) {$\downarrow$};
  \node[spin, right=5mm of L2] (M2) {$\uparrow$};
  \node[spin, right=5mm of M2] (R2) {$\downarrow$};
\end{tikzpicture}

\caption{Edge single spin-flip channels (site $i=1$, OBC) at
$h_0=2J_0$.
(a)~Neighbor down ($s_2=-1$): $|\Delta P_1|=2J_0$.
(b)~Neighbor up ($s_2=+1$): $|\Delta P_1|=6J_0$.
Both channels are absent in PBC where every site has two neighbors.
At $T=2\pi/J_0$, the sinc filter suppresses both edge channels
($a_{nm}=\pi$ and $3\pi$, respectively), and they reappear only at
$\mathcal{O}(g^3)$ through the third-order FPT term~[Eq.~\eqref{eq:HF3_edge_proj}].}
\label{fig:edge_flips}
\end{figure}
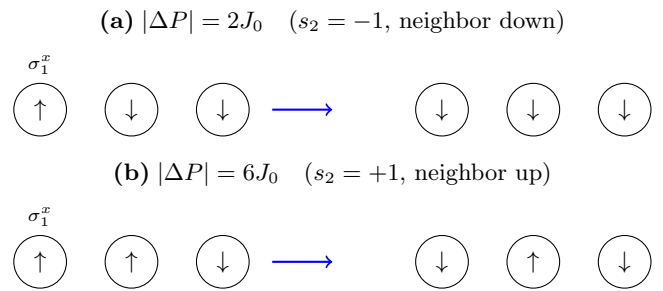
\begin{table}[b]
\caption{Edge single-spin-flip channels for site $i=1$ under OBC at
$h_0=2J_0$.
(a) Neighbor down ($s_2=-1$): $|\Delta P_1|=2J_0$.
(b) Neighbor up ($s_2=+1$): $|\Delta P_1|=6J_0$.
Both channels are absent in PBC, where every site has two neighbors.
At $T=2\pi/J_0$, the sinc filter suppresses both edge channels
($a_{nm}=\pi$ and $3\pi$, respectively). They reappear only at
$\mathcal{O}(g^3)$ through the third-order Floquet perturbation term
[Eq.~\eqref{eq:HF3_edge_proj}].}
\label{tab:sinc_OBC}
\begin{ruledtabular}
\begin{tabular}{lccccc}
$T$ & $|\Delta P|=0$ & $2J_0$ & $4J_0$ & $6J_0$ & $8J_0$ \\
\hline
$\pi/(2J_0)$ & on & on & on & on & off \\
$\pi/J_0$    & on & on & off & on & off \\
$2\pi/J_0$   & on & off & off & off & off \\
\end{tabular}
\end{ruledtabular}
\end{table}

\subsection{Hilbert-space fragmentation}
\label{sec:hsf}
At the primary resonance, the sinc selection rule does more than slow the dynamics. It eliminates all off-resonant single-spin-flip processes and leaves behind only a constrained set of resonant channels. The first-order Floquet generator is therefore not generic but fragmented in its connectivity: it acts nontrivially only within dynamically disconnected subspaces of the full spin-$1/2$ Hilbert space. This is the origin of the prethermal Hilbert-space fragmentation discussed in this section.
\subsubsection{Constrained bulk generator and disconnected sectors}

At the primary resonance $h_0=2J_0$, the drive does \textit{not} freeze the bulk completely. The sinc filter removes all single-spin-flip channels with nonzero change in the diagonal operator $O$, but bulk flips with $\Delta P_i(s)=0$ survive. In the computational basis, this happens precisely when the two neighbors of site $i$ are down. The surviving first-order Floquet generator is a PXP-type kinetic constraint generated \textit{dynamically} inside the full spin-$1/2$ Hilbert space, not imposed by hand. The sinc filter therefore converts a driven unconstrained spin chain into a constrained PXP-type generator at first order \cite{Sala2020PRX,Khemani2020PRBShattering,MoudgalyaMotrunich2022PRX,MoudgalyaBernevigRegnault2022RPP,Ghosh2023PRL,Ghosh2024PRB,Zhang2024StarkHSF}.

We now explore the fragmentation structure. Let $n_i=(\mathbb{I}+\sigma_i^z)/2$ denote the projector onto an up spin on site $i$. The local operator $b_i=n_in_{i+1}$ checks whether the bond $(i,i+1)$ contains an adjacent up-up pair. Since the allowed PXP move flips a spin only when both neighboring sites are down, it can never create or destroy such a pair. Thus
\begin{equation}
b_i=n_in_{i+1},
\qquad
[H_{\mathrm{eff}},b_i]=0
\qquad \forall\, i.
\label{eq:bi_frag_main}
\end{equation}
These conserved bond projectors are not mysterious emergent charges; they simply record which nearest-neighbor up-up bonds are frozen into a basis state. Their binary values partition the Hilbert space into dynamically disconnected Krylov sectors. Choosing non-overlapping bonds already gives an exponential lower bound on the number of sectors, so the fragmentation is extensive rather than accidental \cite{Khemani2020PRBShattering,MoudgalyaMotrunich2022PRX}.

The largest fragment is the sector with no adjacent up spins at all, namely $b_i=0$ on every bond. Its size has a simple physical interpretation. For OBC, the largest \textit{single} fragment is the $(n_1,n_L)=(0,0)$ edge sector and counts binary strings of length $L$ with no adjacent ones and both ends fixed to zero, giving dimension $F_L$. For PBC, the same no-adjacent-up rule must also hold across the boundary, so the largest fragment is a constrained ring of dimension $\mathcal{L}_L=F_{L-1}+F_{L+1}$, the $L$th Lucas number. In both cases the growth is only $\sim \varphi^L$, with $\varphi=(1+\sqrt5)/2<2$, so even the largest fragment occupies an exponentially vanishing fraction of the full Hilbert space: $D_{\max}(L)/2^L\sim (\varphi/2)^L$. The conserved bond projectors therefore partition the Hilbert space into exponentially many disconnected sectors, while the largest sector itself becomes asymptotically negligible compared with $2^L$. That is strong fragmentation \cite{Sala2020PRX,Khemani2020PRBShattering,Rakovszky2020PRBEdgeModes,MoudgalyaBernevigRegnault2022RPP}.

The conserved projectors imply a quantitative lower bound on late-time local memory. For a bulk spin $\sigma_j^z$, we project the local magnetization onto a small conserved operator basis built from the frozen local patterns around site $j$. In Appendix~\ref{app:HSF}, we use three traceless conserved operators: $Q_1$ detects whether the left bond $(j-1,j)$ is an up-up pair, $Q_2$ does the same for the right bond $(j,j+1)$, and $Q_3$ detects the three-site occupancy pattern around $j$. The point of this construction is simple: we want the part of $\sigma_j^z$ that cannot decay under the fragmented dynamics. The Mazur inequality then gives the parameter-independent bound $M_{\sigma_j^z}=3/5$ \cite{Mazur1969Physica,Sirker2020SciPost}.

\subsubsection{Numerical signatures of fragment-restricted thermalization}

\begin{figure}[t]
\centering
\includegraphics[width=\columnwidth]{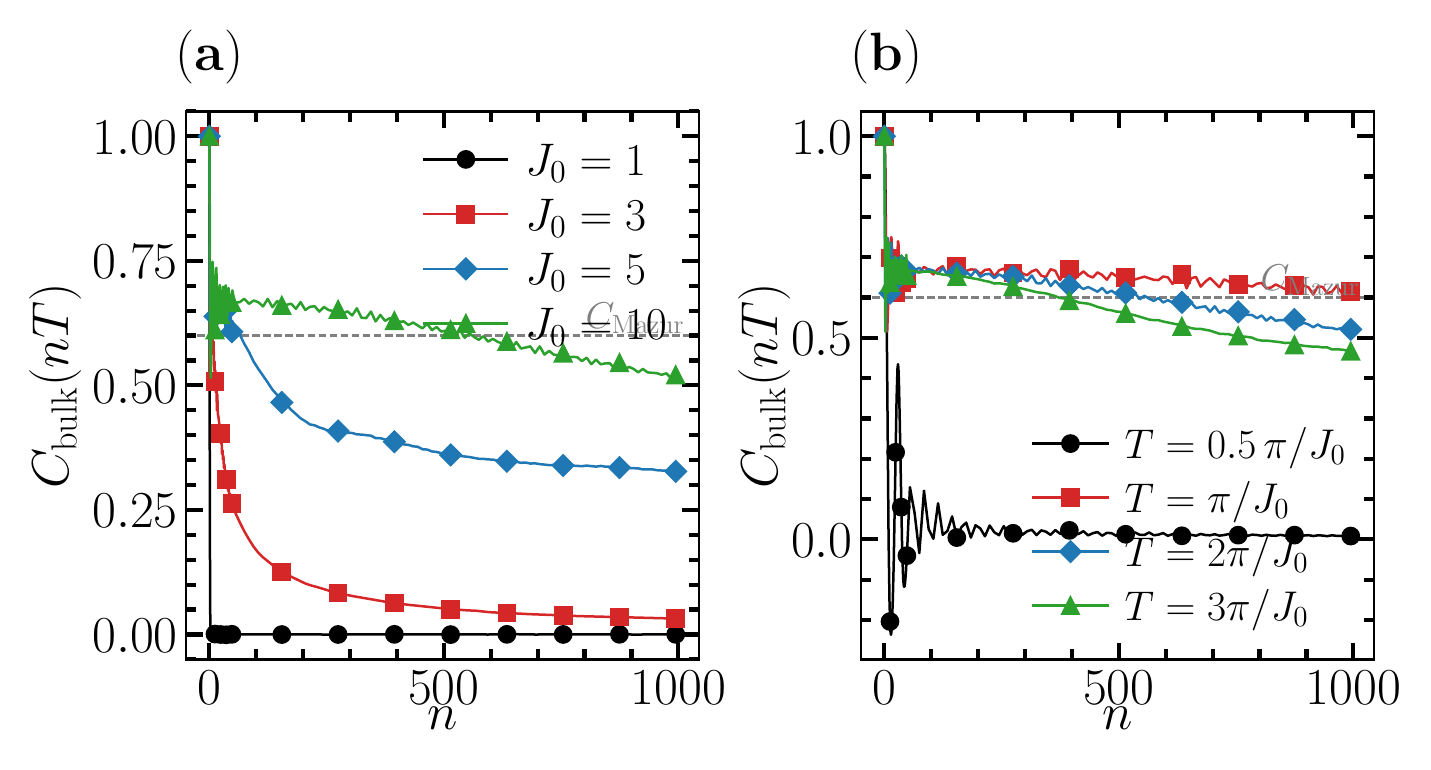}
\caption{
 (a) Plot of the stroboscopic bulk-site autocorrelation, $C_{\mathrm{bulk}}(nT)$ with time (in number of Floquet steps, $n$) at the resonant period $T=2\pi/J_0$ for $J_0\in\{1,3,5,10\}$.  (b) Plot of $C_{\mathrm{bulk}}(nT)$ with at fixed $J_0=10$ and $h_0=20$, with time (in number of Floquet steps, $n$) driven at driving time periods, $T=\mathrm{factor}\times \pi/J_0$ and $\mathrm{factor}\in\{0.5,1,2,3\}$. Both the plots are for a chain of length, $L=16$ and parameters, $g=1$, and $h_0=2J_0$, averaged over $32$ random initial states. The first $50$ Floquet periods are shown at single-period resolution and later times are block-averaged over 10 periods. The grey dashed line marks the Mazur bound $C_{\mathrm{Mazur}}=0.6$.}
\label{fig:auto_bulk}
\end{figure}

\begin{figure}[t]
\centering
\includegraphics[width=\linewidth]{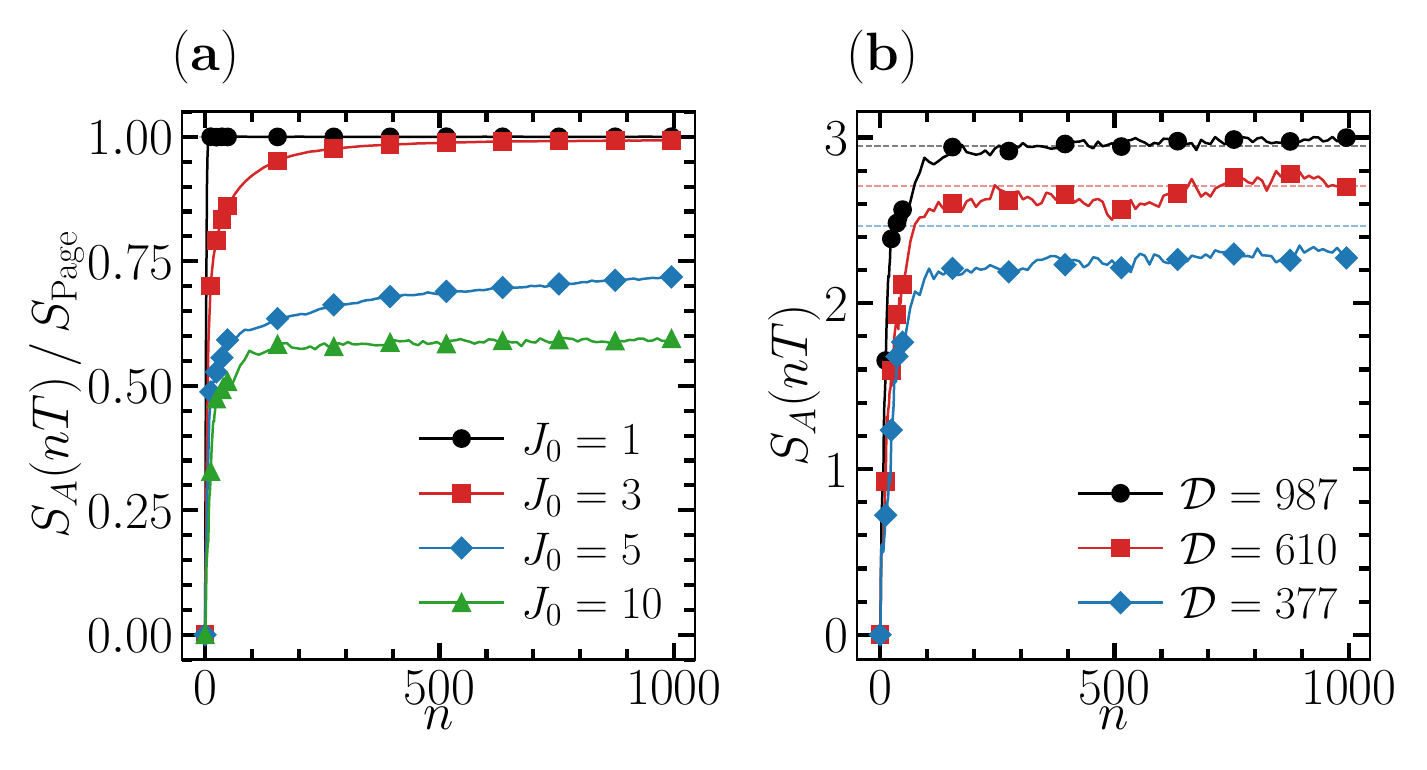}
\caption{
 (a) Stroboscopic half-chain entanglement entropy $S_A(nT)$ plotted with time (as number of Floquet steps, $n$) for the fragmented Floquet chain driven with the resonant period $T=2\pi/J_0$ for $J_0\in\{1,3,5,10\}$, plotted as $S_A/S_{\mathrm{Page}}$ using the Page value of the largest fragment. (b) $S_A(nT)$ plotted with time (as number of Floquet steps, $n$) starting from states from three different Krylov fragments of dimensions $\mathcal{D}=987$, $610$, and $377$ at $J_0=10$, $h_0=20$, and $T=2\pi/J_0$. Dashed lines mark the fragment-specific Page values. Both the plots are for chain of length, $L=16$, with parameters $g=1$, and $h_0=2J_0$, with subsystem $A=\{0,1,\ldots,7\}$. The first 50 Floquet periods are shown at single-period resolution and later times are block-averaged over 10 periods. }
\label{fig:EE_frag}
\end{figure}

\begin{figure}[t]
\centering
\includegraphics[width=\linewidth]{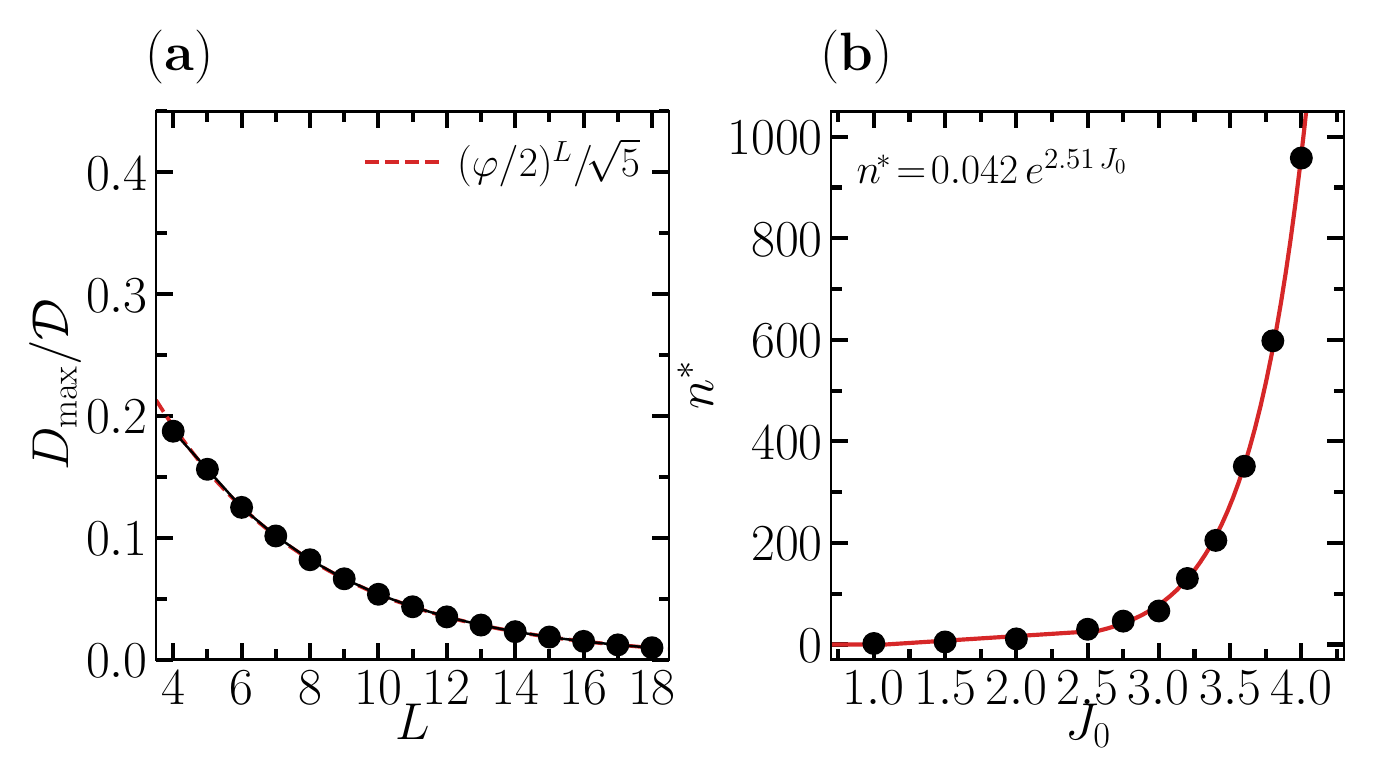}
\caption{
(a) Ratio of the largest fragment dimension $D_{\max}$ to the full Hilbert-space dimension $2^L$ as a function of system size $L$ at the fragmentation resonance. The dashed curve shows the asymptotic Fibonacci scaling $D_{\max}/2^L \sim (\varphi/2)^L/\sqrt{5}$, where $\varphi=(1+\sqrt{5})/2$, demonstrating that even the largest fragment occupies an exponentially vanishing fraction of the full Hilbert space. (b) Bulk-memory decay timescale $n^*(J_0)$ at $L=16$, defined as the first Floquet period after which $|C_{\mathrm{bulk}}(nT)|$ drops below $e^{-1} C_{\mathrm{Mazur}}$ and remains there for five consecutive periods.}
\label{fig:frag_scaling}
\end{figure}

\begin{figure}[t]
\centering
\includegraphics[width=\columnwidth]{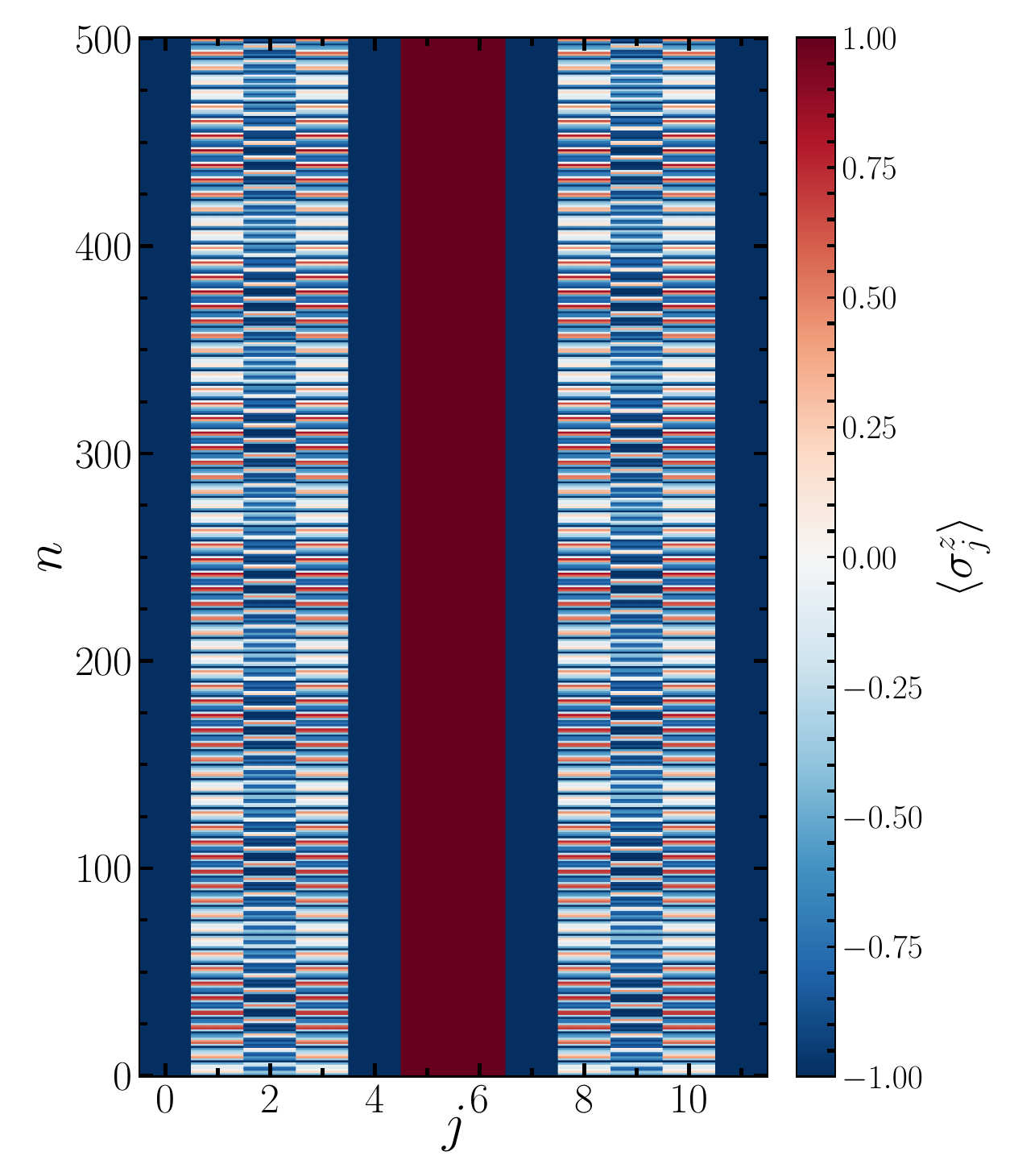}
\caption{\textbf{Real-space confinement of a frozen up-up block.}
Stroboscopic local magnetization $\langle \sigma_j^z(nT)\rangle$ for the Floquet Ising chain at $L=12$, $J_0=10$, $h_0=20$, $g=1$, and $T=2\pi/J_0$, starting from the product state $\ket{\downarrow\downarrow\downarrow\downarrow\downarrow\uparrow\uparrow\downarrow\downarrow\downarrow\downarrow\downarrow}$. The central adjacent up-up block is not melted by the bulk dynamics; instead, the evolution remains confined to the same Krylov sector because the constrained first-order generator cannot create or destroy neighboring up-up pairs. This provides a direct real-space visualization of the conserved bond projectors $b_i=n_i n_{i+1}$.}
\label{fig:frag_realspace}
\end{figure}

Fig.~\ref{fig:auto_bulk} shows that at the resonance point, the bulk autocorrelation develops a long-lived plateau, and increasing $J_0/g$ pushes the system deeper into the prethermal fragmented regime. The plateau remains above the Mazur bound, while detuning the drive period restores rapid decay. The dynamics is slow as the late-time memory is pinned to the existence of local conserved quantities (with respect to the first order Floquet Hamiltonian) and to the same resonance condition that produced Eq.~\eqref{eq:Heff_PXP_OBC}.
Fig.~\ref{fig:EE_frag} addresses the entanglement side of the same story, and here the baseline matters. There are three distinct entropy scales in the problem: the full-space Page value, the fragment-specific Page value, and the observed plateau. These are \textit{not} interchangeable. If the system were ergodic in the full $2^L$ Hilbert space, the half-chain entropy would approach the full-space Page value \cite{Page1993PRL}. Under fragmentation, however, the relevant thermal benchmark is the Page value of the accessible fragment, not that of the full Hilbert space \cite{VidmarRigol2017PRL}. The observed plateau must therefore be read against both scales. In Fig.~\ref{fig:EE_frag}, panel (a), the entropy remains far below the full-space ergodic expectation once the dynamics is deeply prethermal. In Fig.~\ref{fig:EE_frag}, panel (b), the saturation entropy increases systematically with fragment dimension, which is exactly what one expects if the effective thermal volume is the fragment itself. This is the correct distinction between full-space ergodicity and fragment-restricted thermalization. Without it, sub-Page entanglement can be misread as a finite-size artifact; with it, the physics is unambiguous \cite{Deutsch1991PRA,Srednicki1994PRE,RigolDunjkoOlshanii2008Nature,Reimann2008PRL}.
Fig.~\ref{fig:frag_scaling} sharpens the asymptotic and prethermal claims. Fig.~\ref{fig:frag_scaling}, Panel (a) shows that $D_{\max}/2^L$ decays exponentially with system size, in agreement with the Fibonacci/Lucas counting and the asymptotic form $(\varphi/2)^L$. This is the direct numerical signature of strong fragmentation. Fig.~\ref{fig:frag_scaling}, Panel (b) shows that the bulk-memory lifetime grows rapidly with $J_0$. The behavior is consistent with Floquet prethermal fragmentation expectations \cite{Ghosh2023PRL,Kuwahara2016AnnPhys}.
Finally, Fig.~\ref{fig:frag_realspace} gives the real-space picture of the constrained dynamics. Starting from a product state containing a local up-up block, the spatiotemporal magnetization pattern shows that this block remains trapped inside the same Krylov sector. The $(+1,+1)$ motif is not erased by the bulk generator because Eq.~\eqref{eq:bi_frag_main} forbids the creation or destruction of adjacent up-up bonds. The real-space dynamics therefore makes the fragmentation visible without any combinatorics: local patterns are confined by the same constraint that organizes the spectrum.
The sinc filter does not merely slow down the dynamics; at resonance it selects a constrained first-order generator, the resulting conserved bond projectors create exponentially many disconnected sectors, the largest sector is Fibonacci/Lucas-small compared with $2^L$, and the Mazur inequality guarantees a nonzero late-time memory plateau, which is precisely what the stroboscopic numerics display.

\subsection{Prethermal edge memory}
\label{sec:edge}

The boundary behaves differently from the bulk for a microscopic reason: an edge spin has only one neighbor, whereas a bulk spin has two. Because the sinc filter acts on the single-spin-flip energy difference $\Delta P_i$, the set of allowed first-order channels is coordination-sensitive. At the fragmentation resonance $h_0=2J_0$, bulk flips can still satisfy $\Delta P_i=0$ when both neighboring spins are down, which is precisely why the first-order bulk dynamics reduces to the constrained PXP-type generator discussed in Sec.~\ref{sec:hsf}. The edge has no analogous channel. For site $i=1$, the flip cost depends only on the single neighboring spin, so every edge process sits on a nonzero $\Delta P$ channel and is therefore removed by the same sinc filter at first order. This already creates a sharp bulk-edge hierarchy: the bulk remains active within constrained sectors, while the boundary is frozen more strongly.

The lifetime becomes even longer because the symmetric sign-flip protocol also removes the second-order Floquet Hamiltonian contribution. The cancellation arises from the temporal symmetry of the drive: the two halves of the cycle contribute with opposite sign in the second-order perturbation series, producing a time-reversal-like cancellation. As a result,
\begin{equation}
\begin{aligned}
    H_F^{(2)}
&=\frac{1}{2iT}\int_0^T\!dt_1\int_0^{t_1}\!dt_2\;
\bigl[V_I(t_1),\,V_I(t_2)\bigr],\\
H_{F,\mathrm{edge}}^{(2)}&=0.
\end{aligned}
\label{eq:HF2_zero}
\end{equation}
The boundary is therefore not merely weakly coupled at low order; its leading leakage process is pushed all the way to third order.

That third-order process is the first nonvanishing edge-flip channel. At $h_0=2J_0$, the left-edge contribution takes the form
\begin{equation}
\begin{aligned}
H_F^{(3)}
&=\frac{-1}{6T}\!\int_0^T\!dt_1\!\int_0^{t_1}\!dt_2
\!\int_0^{t_2}\!dt_3\;\Bigl(
\bigl[V_I(t_1),\bigl[V_I(t_2),V_I(t_3)\bigr]\bigr]
\\ &+\bigl[V_I(t_3),\bigl[V_I(t_2),V_I(t_1)\bigr]\bigr]
\Bigr),\\
H_{F,\mathrm{edge}}^{(3)}
&=
-\frac{g^3}{2J_0^2}
\left(
\Pi_2^{\downarrow}
+\frac{1}{9}\Pi_2^{\uparrow}
\right)\sigma_1^x,\\
\Pi_2^{\downarrow}&=\frac{\mathbb{I}-\sigma_2^z}{2},
\qquad
\Pi_2^{\uparrow}=\frac{\mathbb{I}+\sigma_2^z}{2}.
\end{aligned}
\label{eq:HF3_edge_proj}
\end{equation}
Here $\Pi_2^{\downarrow}$ and $\Pi_2^{\uparrow}$ project onto the down-spin and up-spin configurations of the site adjacent to the edge. Eq.~\eqref{eq:HF3_edge_proj} makes the physics explicit: the edge is not frozen forever, and it is not protected by topology. The leakage operator exists, but only at order $g^3/J_0^2$. The corresponding edge-memory timescale therefore scales as \begin{equation}
    \tau_{\mathrm{edge}}\sim J_0^2/g^3.
    \label{eq:tau_edge}
\end{equation}

This scaling is the key diagnostic. Because the leading leakage process is local, $\tau_{\mathrm{edge}}$ is independent of the system size $L$. That immediately distinguishes this boundary memory from a strong zero mode or a symmetry-protected edge qubit. There is no exponentially long lifetime in $L$, no topological protection, and no exact asymptotic edge conservation. What survives instead is a \textit{prethermal} boundary polarization: algebraically long-lived in the control parameter $J_0/g$, but not stable in the thermodynamic sense associated with genuine strong zero modes or Floquet symmetry-protected edge phases \cite{ElseFendleyKempNayak2017PRX,KempYaoLaumann2020PRL,ElseNayak2016PRB,VonKeyserlingkSondhi2016PRB1,PotterMorimotoVishwanath2016PRX,HarperRoyRudnerSondhi2020ARCMP}. In that sense, the edge phenomenon here is perturbatively generated prethermal edge memory as compared to a topological edge mode \cite{ParkerVasseurScaffidi2019PRL,YatesMitra2022CommunPhys,SurSen2024JPCM,Rakovszky2020PRBEdgeModes}.

\begin{figure}[t]
\centering
\includegraphics[width=\columnwidth]{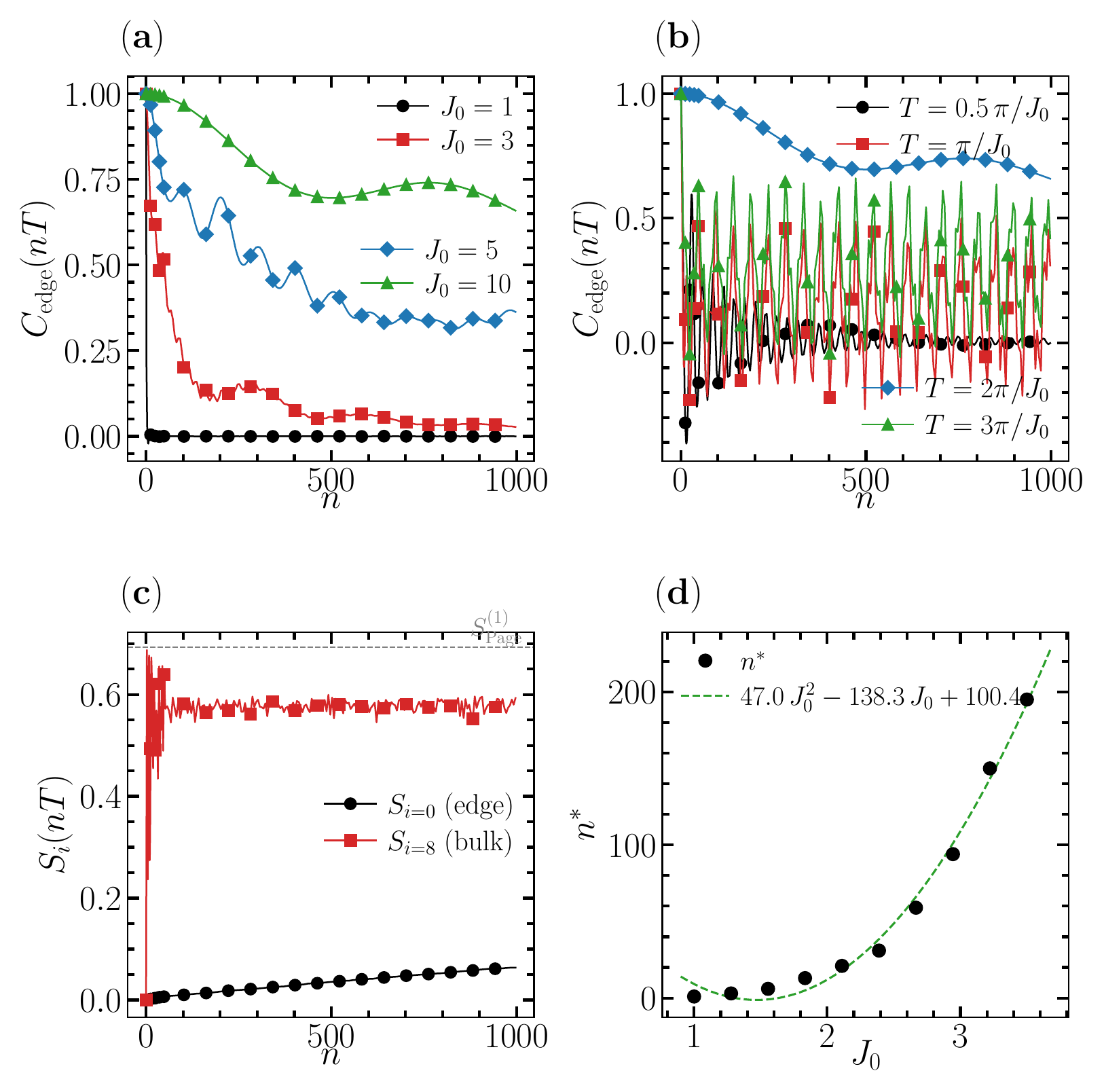}
\caption{
(a) Autocorrelation at the edge site, $C_{\mathrm{edge}}(nT)$ plotted with time (as number of Floquet steps, $n$) for $J_0\in\{1,3,5,10\}$ at $T=2\pi/J_0$. Increasing $J_0/g$ suppresses the leading leakage channel and enhances the long-lived boundary polarization. (b) $C_{\mathrm{edge}}(nT)$ for fixed $J_0=10$ and $h_0=20$, with $T=\mathrm{factor}\times \pi/J_0$ and $\mathrm{factor}\in\{0.5,1,2,3\}$. (c) Single-site vs rest of the chain partition of entanglement entropy for an edge site and a bulk site at $J_0=10$ and $T=2\pi/J_0$. (d) Extracted edge-decay time $n^\ast(J_0)$ versus $J_0$. The growth is consistent with the perturbative expectation $\tau_{\mathrm{edge}}\sim J_0^2/g^3$ for larger $J_0$ values, indicating parametrically long but non-topological edge memory. Open-chain edge diagnostics at the resonance point $h_0=2J_0$. Autocorrelation panels are averaged over $32$ random initial states; The entanglement entropy plots start the dynamics from a random state in the largest fragment. The first 50 Floquet periods are shown at single-period resolution and later times are block-averaged over 5 periods for all the plots.}
\label{fig:auto_edge}
\end{figure}

Fig.~\ref{fig:auto_edge} shows this hierarchy directly. Fig.~\ref{fig:auto_edge}, Panel (a) gives the coupling scan: increasing $J_0/g$ suppresses the third-order leakage scale and leaves the edge polarization coherent for longer times. Fig.~\ref{fig:auto_edge}, Panel (b) shows that the effect is sharply period-selective. The strongest boundary memory occurs at the same resonant period that generates the fragmented bulk Hamiltonian, which rules out that this is just some generic boundary artifact. The prethermal oscillations at $T=\{1,3\}\times \pi/J_0$ are due to the partial destructive interference which allows for edge flips but in a restrictive sense as the bulk flips are still  repressed~\cite{GhoshPaulSengupta2025Spin}.  Fig.~\ref{fig:auto_edge}, Panel (c) provides the entanglement counterpart: the bulk site entangles rapidly, while the edge-site entropy remains strongly suppressed, exactly as expected when the boundary leakage operator is pushed to third order. Fig.~\ref{fig:auto_edge}, Panel (d) connects the numerics back to perturbation theory. The extracted edge-decay time grows consistently with the expectation $\tau_{\mathrm{edge}}\sim J_0^2/g^3$.
The conclusion is therefore precise. The Floquet filter generates a coordination-selective perturbative hierarchy: first-order constrained dynamics survives in the bulk, while the edge is suppressed through second order and leaks only at third order. The result is long-lived, non-topological boundary polarization prethermal edge memory which is not a strong zero mode ($L$ independent).

\subsection{Largest-fragment PXP scars}
\label{sec:scars}

Under periodic boundary conditions, the boundary-specific channel suppression discussed in the edge analysis is absent. At the commensurate drive where the first-order selection rule leaves only the no-adjacent-up sector active, the largest fragment is therefore not merely PXP-like: it is exactly the constrained PXP Hilbert space at first order. The relevant question is then not whether PXP scars exist as they do, but whether the Floquet dynamics inherits that structure within the prethermal window before higher-order terms restore leakage beyond the first-order description.

\begin{equation}
H_{\mathrm{eff}}\big|_{\mathcal{H}^{\mathrm{PBC}}_{\max}} = H_{\mathrm{PXP}},
\label{eq:Heff_PXP_restricted}
\end{equation}
where $\mathcal{H}^{\mathrm{PBC}}_{\max}$ is the largest periodic-boundary fragment and $H_{\mathrm{PXP}}$ is the standard constrained PXP Hamiltonian acting on the no-adjacent-up subspace. It follows that the appropriate diagnostics are the standard PXP scar diagnostics restricted to this fragment: low-entanglement outliers in the Floquet spectrum, anomalously large overlap with the charge-density-wave (CDW) state
\[
|\mathbb{Z}_2\rangle = |\uparrow\downarrow\uparrow\downarrow\cdots\rangle,
\]
non-random participation count $N_{\mathrm{eff}}$, sub-GOE or otherwise non-fully-ergodic level statistics inside symmetry-resolved sub-blocks, and coherent revivals from $|\mathbb{Z}_2\rangle$~\cite{FendleySenguptaSachdev2004PRB,Bernien2017Nature,Turner2018NatPhys,Mukherjee2020PRBFloquetScars,MizutaTakasanKawakami2020PRResearch,SugiuraKuwaharaSaito2021PRResearch,HudomalDesaulesMukherjeeSuHalimehPapic2022PRB,Giudici2024PRL,PaiPretko2019PRL,Ghosh2026arXiv}.

\begin{figure}[t]
    \centering
    \includegraphics[width=1.1\linewidth, height=0.6\linewidth]{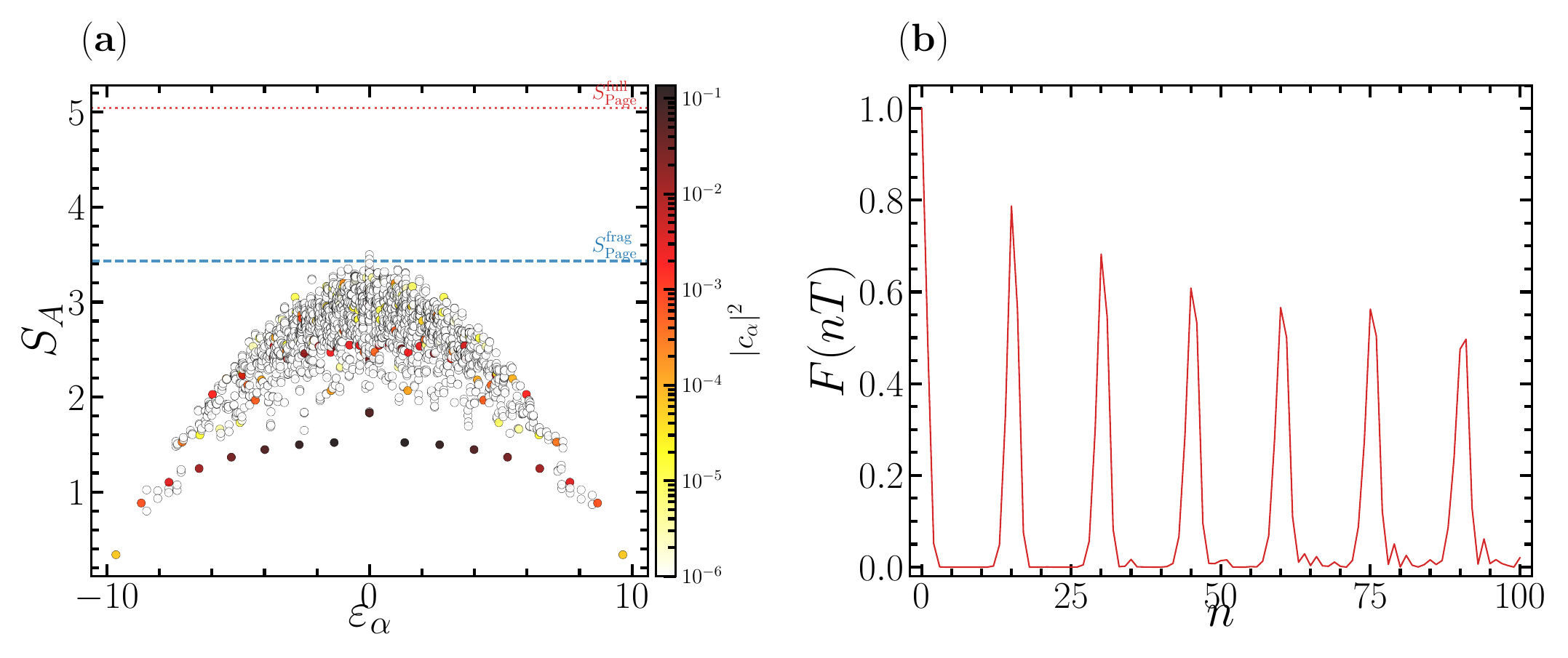}
    \caption{(a) Floquet eigenstate scatter for the largest Fibonacci fragment($D = 2207$) of the periodically driven Ising chain at $L = 16$, $J_0 = 10.0$, $h_0 = 2\,J_0 = 20$, $g = 1.0$, $T = \pi/J_0$.  Each point is a Floquet eigenstate $|\phi_\alpha\rangle$; $z$-axis (colour axis) shows the $\mathbb{Z}_2$ CDW overlap $|c_\alpha|^2 =|\langle \mathbb{Z}_2|\phi_\alpha\rangle|^2$ (log scale), horizontal axis is the quasienergy $\varepsilon_\alpha$, and the vertical axis encodes the half-chain entanglement entropy $S_A$.  Dashed vertical line marks $1/D$ (uniform weight).  States with $|c_\alpha|^2 \gg 1/D$ and low $S_A$ (dark colour) are scar candidates: their anomalously large $\mathbb{Z}_2$ overlap and sub-Page entanglement indicate non-ergodic Floquet eigenstates.  Blue/red lines on the colourbar mark the fragment/full-space Page values.
    (b) Stroboscopic return fidelity $F(nT) = |\langle \mathbb{Z}_2|U_F^n|\mathbb{Z}_2\rangle|^2$
    for the $\mathbb{Z}_2$ CDW initial state at $L = 14$, same Hamiltonian
    parameters.  Dashed blue line: inverse fragment dimension $1/D_{\mathrm{frag}}$;
    dotted red line: ergodic floor $1/2^L$.  Fragmentation confines dynamics
    to the Fibonacci sector, so $F(nT)$ saturates above $1/D_{\mathrm{frag}}$ rather
    than decaying to $1/2^L$.  Partial revivals reflect coherent oscillations
    within the fragment.}
    \label{fig:fig7}
\end{figure}

The eigenstate scatter in Fig.~\ref{fig:fig7}, Panel (a) provides the spectral evidence. Writing
$|c_\alpha|^2 := |\langle \mathbb{Z}_2|\phi_\alpha\rangle|^2$, where
$|\phi_\alpha\rangle$ is a Floquet eigenstate, scar candidates are precisely those states with
$|c_\alpha|^2 \gg 1/D_{\mathrm{frag}}$ and anomalously low half-chain entanglement entropy $S_A$, with
$D_{\mathrm{frag}}$ the dimension of the largest fragment.The crucial comparison is to the \textit{fragment} Page value, not the full-space Page value. In this sense, the fragment matters spectrally as well as dynamically.

The corresponding dynamics, shown in Fig.~\ref{fig:fig7}, Panel (b), is captured by the return fidelity
\begin{equation}
F(n)=\left|\langle \mathbb{Z}_2|U_F^n|\mathbb{Z}_2\rangle\right|^2,
\label{eq:Z2_return_fidelity}
\end{equation}
where $U_F$ is the one-period Floquet unitary and $n$ is the number of drive periods. Pronounced revivals of $F(n)$ indicate that $|\mathbb{Z}_2\rangle$ has atypically large weight on a small set of special Floquet eigenstates rather than spreading ergodically across the fragment. The correct ergodic floor is therefore $1/D_{\mathrm{frag}}$, not $1/2^L$: once fragmentation confines the dynamics to a single sector, the full Hilbert-space baseline is physically irrelevant. The fact that the revival peaks remain parametrically above $1/D_{\mathrm{frag}}$ is exactly the dynamical signature expected from inherited PXP scar structure.

Structurally, this is the same logic used in recent driven scar problems: the scar regime is identified not by revivals alone, but by the specific drive window in which the first-order Floquet generator takes a scar-hosting form~\cite{Ghosh2026arXiv}. In our case, at the resonant commensurate drive, the largest fragment becomes PXP at first order, and the Floquet dynamics inherits the corresponding scar phenomenology inside that fragment. Higher-order terms will eventually spoil exact PXP dynamics, so the statement is explicitly prethermal. The scar sector is not imposed by hand but is generated dynamically by the sinc-based selection rule that produced the fragmentation in the first place.
\subsection{Floquet freezing}
\label{sec:freezing}

Floquet freezing in this model is the complementary outcome of the same first-order selection rule that produced Hilbert-space fragmentation. The distinction is sharp and should not be blurred. When the drive parameters leave some single-spin-flip channels resonant, the first-order Floquet generator remains nontrivial and organizes the dynamics into disconnected sectors. When the same sinc filter suppresses \textit{all} such channels, there is no residual constrained motion left at order $g$: the system is frozen to first order.

This difference can be summarized at the level of the local single-flip eigenvalue difference $\Delta P$. At the fragmentation point,
\begin{equation}
\begin{aligned}
   h_0&=2J_0
\\
 \Longrightarrow \exists\, \Delta P &= 0 \text{ channels}
\Longrightarrow 
\text{fragmentation}, 
\end{aligned}
\label{eq:fragmentation_vs_freezing_a}
\end{equation}
so some first-order processes survive and generate nontrivial in-fragment dynamics. By contrast, for the higher commensurate ratios
\begin{equation}
\begin{aligned}
    h_0&=2nJ_0,~ n>1 \\
\Longrightarrow 
\Delta P &\neq 0 \text{ for all single-flip channels}
\Longrightarrow
H_F^{(1)}=0,
\end{aligned}
\label{eq:fragmentation_vs_freezing_b}
\end{equation}
with $n$ a positive integer labeling the commensurate field ratio and $H_F^{(1)}$ the first-order Floquet Hamiltonian. This is the freezing line. It is not a fragmented regime with slower motion; it is a regime in which the leading generator of motion vanishes altogether.

 In the HSF regime, $H_F^{(1)}$ is constrained but active: the Hilbert space breaks into disconnected sectors, and the system can still evolve nontrivially within a given sector. In the freezing regime, $H_F^{(1)}=0$, so the system barely moves at leading order. Put differently, fragmentation retains structure in the dynamics, whereas freezing removes it.

\begin{figure}[t]
\centering
\includegraphics[width=\columnwidth]{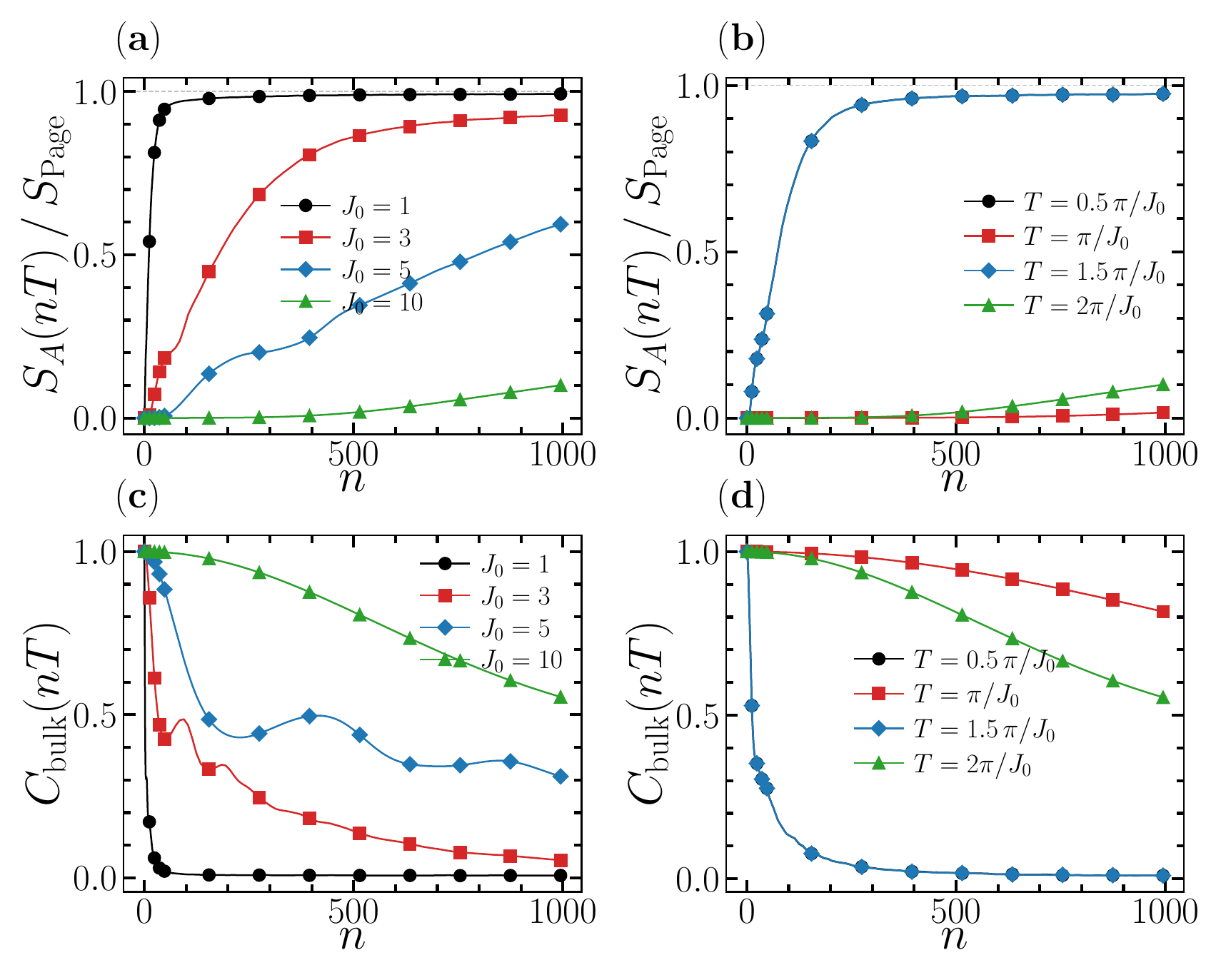}
\caption{
Top row: half-chain entanglement entropy normalized by the Page value. Bottom row: bulk autocorrelation. Panel (a,c): coupling scan at fixed $T=2\pi/J_0$. Panel (b,d): period scan at fixed $J_0=10$. The freezing point suppresses both entanglement growth and relaxation of the bulk memory.}
\label{fig:freezing_mag}
\end{figure}

The numerics in Fig.~\ref{fig:freezing_mag} are therefore confirmatory this. The top row shows strongly suppressed growth of the half-chain entanglement entropy, meaning that the dynamics fails to efficiently explore Hilbert space. The bottom row shows a persistent bulk autocorrelation plateau, meaning that local memory is retained for anomalously long times. The scan over time periods of autocorrelation and entanglement entropy is especially important because it rules out a trivial large-$J_0$ explanation. At fixed large $J_0$, the strongest suppression occurs at the special commensurate period $T=2\pi/J_0$, where the first-order spin-flips are disabled. Nearby periods permit residual first-order processes and therefore faster entropy growth and faster decay of local memory. The effect is thus tied to commensurate Floquet engineering, not merely to a large bare scale.

For open chains there is also a weaker, incomplete version of this effect at $T=\pi/J_0$. There the bulk channels are suppressed, but some boundary channels remain active, so the system is only partially frozen. This is incomplete channel suppression: the bulk memory is protected more strongly than the boundary, and the residual edge dynamics prevents true first-order arrest. Thus, fragmentation and freezing are not competing explanations of the dynamics; they are the two different outcomes of the same selection rule. If some resonant first-order channels survive, the result is a constrained active Hamiltonian and fragmented dynamics. If none survive, the result is first-order Floquet freezing. 

\subsection{Generalisation of the driven Ising model to higher dimensions and different lattice geometries}
\label{sec:graph_generalization}

The mechanism is not one-dimensional. Let $G=(V,E)$ be a finite induced subgraph of a $z$-regular parent graph, and let $d_i$ denote the degree of site $i$ within $G$. Consider the driven Ising model
\begin{equation}
H(t)=-\lambda(t)\left[J\sum_{(ij)\in E}Z_iZ_j+h\sum_{i\in V}Z_i\right]-g\sum_{i\in V}X_i,
\label{eq:graph_model}
\end{equation}
with the same symmetric sign-flip protocol $\lambda(t)=\pm1$. Because the diagonal part again closes over one period, the first-order Floquet Hamiltonian inherits the sinc filter as in the paper.

For a single spin flip at site $i$, the relevant gap is now
\begin{equation}
\Delta_i(s)=-2s_i\left(J\sum_{j\in\partial i}s_j+h\right),
\label{eq:Delta_i_graph}
\end{equation}
where $\partial i$ is the neighbor set of site $i$ in $G$. Choosing $h=zJ$ and the commensurate period $T_\star=2\pi/J$, one obtains a sharp coordination-selective rule. Bulk sites with $d_i=z$ admit a zero-gap channel only when all their neighbors are down, whereas boundary sites with $d_i<z$ cannot satisfy $\Delta_i=0$ at all. The first-order Floquet Hamiltonian is therefore
\begin{equation}
H_F^{(1)}(T_\star)=
-g\sum_{i:\,d_i=z}\left(\prod_{j\in\partial i}\pi_j\right)X_i,
\label{eq:HF1_graph_final}
\end{equation}
which is the graph analogue of the one-dimensional constrained Hamiltonian. In plain language: the same Floquet filter that isolates the bulk for flips in the open chain in one dimension isolates the full-coordination core of any finite piece of a regular Ising graph. 

\section{Discussion and conclusion}
\label{sec:discussion}

The central result of this work is an analytic result for engineering constrained Floquet dynamics in a driven spin-$1/2$ chain. The key object is a sinc-based interference filter that acts on local single-spin-flip processes in the first-order Floquet expansion. That single filter produces three distinct outcomes. When a subset of first-order channels survives in a constrained form, the effective generator $H_F^{(1)}$ remains nontrivial and the dynamics fragments into disconnected sectors. When all first-order channels are suppressed, $H_F^{(1)}$ vanishes and the system exhibits Floquet freezing. Under open boundary conditions, the reduced coordination at the boundary modifies the same channel structure, and in combination with higher-order leakage this produces long-lived edge memory. Fragmentation, freezing, and edge retention are therefore not separate mechanisms in our model, but three consequences of the same analytic selection rule.

This places the model in a useful position relative to earlier Floquet-fragmentation constructions. Compared with recent driven fermionic settings, the gain here is analytic transparency in the full spin-$1/2$ Hilbert space: the surviving and suppressed processes can be identified directly from local energy differences, the fragmented sectors can be counted explicitly through Fibonacci and Lucas combinatorics, and the boundary effect is tied to the same microscopic selection rule rather than introduced as an unrelated feature~\cite{Ghosh2023PRL,Ghosh2024PRB,GhoshPaulSengupta2025Spin}. At the same time, the present mechanism should not be confused with exact Hilbert-space fragmentation in static systems. Here the constraints are prethermal rather than exact: they arise from the leading Floquet generator and are eventually violated by higher-order processes~\cite{Sala2020PRX,Khemani2020PRBShattering}.

The boundary memory requires the same level of precision. It is not a strong-zero-mode effect and not a topological edge-coherence mechanism. The boundary spin remains long-lived because the first-order boundary channels are suppressed more strongly than the bulk channels, while higher-order terms generate leakage only perturbatively. The resulting lifetime is therefore structural rather than symmetry-protected, with the scaling $\tau_{\mathrm{edge}} \sim \frac{J_0^2}{g^3},$ where $\tau_{\mathrm{edge}}$ is the edge-memory lifetime, $J_0$ is the Ising coupling scale, and $g$ is the transverse-spin-flip amplitude. Its algebraic dependence on $g$ and weak dependence on system size sharply distinguish it from topological or strong-zero-mode edge memory~\cite{ElseFendleyKempNayak2017PRX,KempYaoLaumann2020PRL,Mukherjee2024PRBEdgeMemory}.

The largest-fragment scar phenomenology should also be interpreted carefully. We do not claim to have constructed a new intrinsic scar Hamiltonian. Rather, the point is that the same Floquet selection rule dynamically engineers the largest periodic fragment so that its first-order generator is exactly the PXP Hamiltonian. The scar sector is therefore not imposed by hand; it emerges as the largest fragment of the same Floquet mechanism that also yields fragmentation and freezing. The resulting low-entanglement outliers, enhanced $\mathbb{Z}_2$ overlap, and coherent revivals should accordingly be understood as inherited PXP-like scar physics within the prethermal window, not as evidence for exact scars of the full Floquet unitary~\cite{Turner2018NatPhys,Mukherjee2020PRBFloquetScars,Ghosh2026arXiv}.

The ingredients of the model are Ising interactions, longitudinal fields, transverse flips, and square-wave modulation which are standard in detuned dressed-Rydberg and trapped-ion platforms, suggesting that the predicted signatures may be accessible in programmable driven spin simulators~\cite{Bernien2017Nature,BluvsteinEtAl2021Science,BlochDalibardZwerger2008RMP,TarruellSanchezPalencia2018CRPhys}.

In summary, the paper identifies a single analytic Floquet filter and shows that it organizes the driven dynamics into three regimes: constrained first-order motion and Hilbert-space fragmentation, complete first-order suppression and Floquet freezing, and boundary-selective suppression with perturbative edge memory. The same mechanism also dynamically generates the largest-fragment PXP sector. That combination of analytic control, fragment counting, edge physics, and Floquet-engineered scar phenomenology is the real content of the work.

\begin{acknowledgments}
We thank Biswajit Paul for useful discussions. T.M.\ acknowledges support from
Science and Engineering Research Board (SERB), Govt., through project No. MTR/2022/000382 and
STR/2022/000023. R.P.J.\ acknowledges support from INSPIRE fellowship (Department of Science and Technology, Government of India). SNM acknowledges support from the Kishore Vaigyanik Protsahan Yojana (KVPY) fellowship, SX-2011055, awarded by the Department of Science and Technology, Government of India. Numerical calculations were performed on the high-performance computing facility at the National Institute of Science Education and Research (NISER), Jatni.
\end{acknowledgments}

\appendix

\section{First-order sinc filter}
\label{app:floquet}

This appendix explains why the symmetric sign-flip protocol collapses the first-order interaction picture integral into a single half-cycle contribution and therefore produces the sinc envelope in Eq.~\eqref{eq:HF1_sinc_main}.

Sinnce the diagonal part $H_0(t)$ reverses sign exactly at $t=T/2$, the
interaction-picture perturbation has the piecewise form
\begin{equation}
V_I(t)=
\begin{cases}
e^{-i\mathcal{O}t}\,H_1\,e^{i\mathcal{O}t}, & 0\le t\le T/2,\\[4pt]
e^{-i\mathcal{O}(T-t)}\,H_1\,e^{i\mathcal{O}(T-t)}, & T/2<t\le T.
\end{cases}
\label{eq:app:VI_piecewise}
\end{equation}
Here $\mathcal{O}$ is the diagonal operator defined in the main text, and
$H_1=-g\sum_i \sigma_i^x$ is the transverse perturbation. The second line
is the first line with $t\to T-t$. This is the only structural fact that
matters: the second half of the drive retraces the first half in the
interaction picture.

As a result, the first-order Dyson term is not the sum of two unrelated
integrals. The two half-cycles are identical after the change of
variables $\tau=T-t$, so
\begin{equation}
U_1^I(T,0)
=
-i\int_0^T dt\,V_I(t)
=
-2i\int_0^{T/2} dt\;
e^{-i\mathcal{O}t}\,H_1\,e^{i\mathcal{O}t}.
\label{eq:app:U1I_doubled}
\end{equation}
This exact doubling is what produces the finite-time interference factor.
It is special to the symmetric sign-flip protocol and would not survive
for a generic asymmetric drive.

Now evaluate Eq.~\eqref{eq:app:U1I_doubled} in the eigenbasis of
$\mathcal{O}$, defined by $\mathcal{O}|m\rangle=P_m|m\rangle$, where
$P_m$ is the $\mathcal{O}$-eigenvalue of $|m\rangle$. Writing
$P_{nm}:=P_n-P_m$, one finds
\begin{equation}
\bigl(U_1^I\bigr)_{nm}
=
-iT\,(H_1)_{nm}\,
\operatorname{sinc}\!\left(\frac{P_{nm}T}{4}\right)
e^{-iP_{nm}T/4}.
\label{eq:app:U1I_sinc}
\end{equation}
Comparing
$U_F=\mathbb{I}+U_1^I+\mathcal{O}(g^2)=e^{-iH_FT}$
immediately gives the main-text selection rule,
Eq.~\eqref{eq:HF1_sinc_main}.

The physical interpretation is simple. A single-spin-flip channel
accumulates a phase set by the $\mathcal{O}$-eigenvalue difference
$P_{nm}$ during the first half-cycle, and the second half-cycle partially
undoes it. The residual amplitude after one period is therefore the sinc
interference factor. Channels with $P_{nm}=0$ survive, while channels
with $P_{nm}T/4=\pi k$ for nonzero integer $k$ are removed at first
order. Since $H_1$ connects computational-basis states only by one spin
flip, the entire first-order Floquet problem reduces to classifying which
local single-flip processes satisfy those conditions.
\section{Allowed single-flip $\Delta P$ values for OBC and PBC}
\label{app:Pnm}

At the resonance $h_0=2J_0$, the first-order Floquet selection rule is
controlled entirely by the $\mathcal{O}$-eigenvalue difference produced by
a \textit{single} spin flip. It is therefore unnecessary to enumerate the full
$\mathcal{O}$ spectrum of a finite chain. The allowed $\Delta P$ values are
fixed locally by the number and orientation of the neighbouring spins.

For open boundary conditions, bulk flips and edge flips are inequivalent.
A bulk spin has two neighbours and produces the same three channels as in the
main text, while an edge spin has only one neighbour and therefore generates
two additional values. The distinct magnitudes are
\begin{equation}
|\Delta P|_{\mathrm{OBC}}
\in
\{0,\;2J_0,\;4J_0,\;6J_0,\;8J_0\}.
\label{eq:app:DP_obc_set}
\end{equation}
The values $2J_0$ and $6J_0$ are edge-only channels, whereas
$0$, $4J_0$, and $8J_0$ already occur in the bulk.

For periodic boundary conditions, every site has two neighbours, so the
edge-only channels are absent. The allowed set reduces to
\begin{equation}
|\Delta P|_{\mathrm{PBC}}
\in
\{0,\;4J_0,\;8J_0\}.
\label{eq:app:DP_pbc_set}
\end{equation}
This is the only structural difference between OBC and PBC at first order,
but it has an important consequence for the Floquet filtering.

Because the sinc factor vanishes when $|\Delta P|\,T/4=\pi k$ with nonzero
integer $k$, periodic chains reach the PXP-filtered regime at a shorter
period than open chains:
\begin{equation}
T_{\mathrm{PBC}}^{\ast}=\frac{\pi}{J_0},
\qquad
T_{\mathrm{OBC}}^{\ast}=\frac{2\pi}{J_0}.
\label{eq:app:period_compare}
\end{equation}
The reason is simple: in PBC the nonzero channels are only $4J_0$ and $8J_0$,
both of which already lie on sinc zeros at $T=\pi/J_0$. In OBC the extra
edge channels $2J_0$ and $6J_0$ survive at that shorter period and are removed
only at $T=2\pi/J_0$.

The same reasoning applies for any $L\ge 3$. Since $\Delta P$ is determined
only by the local neighbourhood of the flipped spin, the sets
in Eqs.~\eqref{eq:app:DP_obc_set} and \eqref{eq:app:DP_pbc_set} are
independent of system size.

\section{From the zero-gap condition to the PXP projector structure}
\label{app:PXP}

At $h_0=2J_0$ and $T=2\pi/J_0$ for OBC, the sinc filter removes every
first-order channel except those with $\Delta P_i(s)=0$. Since the
perturbation $H_1=-g\sum_i \sigma_i^x$ connects computational-basis
states only by single-spin flips, the derivation reduces to a local
question: for which spin configurations does flipping site $i$ leave the
$\mathcal O$-eigenvalue unchanged?

For a bulk site, the zero-gap condition is equivalent to requiring both
neighbours of the flipped spin to be down. In the computational basis this
can be written as
\begin{equation}
\delta_{\Delta P_i(s),\,0}
=
\delta_{s_{i-1},-1}\,\delta_{s_{i+1},-1},
\qquad 2\le i\le L-1.
\label{eq:app:delta_constraint_short}
\end{equation}
Thus a bulk spin can flip at first order only when it is flanked by two
down spins. By contrast, an edge spin has only one neighbour, and its
local $\Delta P$ never vanishes at this resonance. Therefore edge flips
are excluded completely.

To express this condition as an operator, introduce the projector onto a
down spin,
\begin{equation}
\pi_i\ket{s}
=
\frac{1-s_i}{2}\ket{s},
\label{eq:app:pi_action_short}
\end{equation}
so that \(\pi_i\) acts as the identity when \(s_i=-1\) and annihilates
the state when \(s_i=+1\). Then the constrained flip operator has matrix
elements
\begin{equation}
\bra{s'}\pi_{i-1}\sigma_i^x\pi_{i+1}\ket{s}
=
\delta_{s',\,s^{(i)}}
\,\delta_{s_{i-1},-1}\,\delta_{s_{i+1},-1},
\label{eq:app:pxp_me_short}
\end{equation}
where \(\ket{s^{(i)}}\) denotes the configuration obtained from
\(\ket{s}\) by flipping spin \(i\).

Equation~\eqref{eq:app:pxp_me_short} matches exactly the surviving
first-order matrix elements selected by
Eq.~\eqref{eq:app:delta_constraint_short}. Therefore the filtered
Floquet generator is precisely the constrained PXP Hamiltonian quoted in
the main text, Eq.~\eqref{eq:Heff_PXP_OBC}. The logic is simple: the
Floquet filter first enforces \(\Delta P=0\), and at the resonance
\(h_0=2J_0\) that condition is equivalent to the local kinetic constraint
that both neighbouring spins must be down before a flip is allowed.

For periodic boundary conditions the same argument applies at every site,
because there are no edges. The only difference is that the constrained
sum runs over all \(i\), and the required period is shorter,
\(T=\pi/J_0\), since the edge-only channels are absent.

\section{Fragmentation details: largest sector and Mazur projection}
\label{app:HSF}

The local conservation law
\(
b_i=n_i n_{i+1}
\)
and the physical reason for
\(
[H_{\mathrm{eff}},b_i]=0
\)
were already explained in Sec.~\ref{sec:hsf}: an allowed PXP flip acts only when both neighbours are down, so it cannot create or destroy an adjacent up-up bond. We therefore keep only the points not shown explicitly in the main text.

\subsection{Exponential sector count and the largest fragment}
\label{app:HSF:largest}

Choosing non-overlapping bonds, for example
\(
\{b_2,b_4,b_6,\dots\},
\)
already gives
\begin{equation}
N_{\mathrm{sectors}}\ge 2^{\lfloor (L-1)/2\rfloor},
\label{eq:app:sector_lower_bound}
\end{equation}
since each such bond can independently take the values \(0\) or \(1\), and sectors with different bond patterns cannot be connected by \(H_{\mathrm{eff}}\).

To see why the no-adjacent-up sector is the largest, consider any sector with at least one bond satisfying \(b_i=1\). Then the configuration contains a maximal block of \(m\ge 2\) consecutive up spins. Under the constrained generator, that block and its two neighbouring sites are frozen, because a spin can flip only if both of its neighbours are down. Hence at least \(m+2\ge 4\) sites are removed from the active Hilbert space.

For PBC, the remaining active region has length at most \(L-m-2\), so its dimension is bounded by
\begin{equation}
D_{\mathrm{sector}}^{\mathrm{PBC}}
\le F_{L-m}
\le F_{L-2}
< \mathcal{L}_L ,
\label{eq:app:pbc_largest_bound}
\end{equation}
where \(F_\ell\) is the Fibonacci number and \(\mathcal{L}_L\) is the Lucas number. Therefore the no-adjacent-up ring, of dimension \(\mathcal{L}_L\), is strictly the largest PBC fragment.

For OBC, any sector with \(n_1=1\) or \(n_L=1\) already freezes one extra site at the boundary and has dimension at most \(F_{L-1}\). If instead \(n_1=n_L=0\) but some internal bond has \(b_i=1\), the same frozen-block argument gives a bound \(\le F_{L-2}\). Hence the \((n_1,n_L)=(0,0)\), \(\mathbf{b}=\mathbf{0}\) sector, whose dimension is \(F_L\), is strictly the largest OBC fragment:
\begin{equation}
F_L > F_{L-1} > F_{L-2}\qquad (L\ge 4).
\label{eq:app:obc_largest_bound}
\end{equation}

\subsection{Mazur bound for bulk memory}
\label{app:HSF:mazur}

For a non-orthogonal conserved set \(\{Q_\alpha\}\), the Mazur bound for an observable \(A\) is
\begin{equation}
M_A=\mathbf{v}^{\,T}K^{-1}\mathbf{v},
\qquad
v_\alpha=\frac{\mathrm{Tr}(A Q_\alpha)}{D},
\qquad
K_{\alpha\beta}=\frac{\mathrm{Tr}(Q_\alpha Q_\beta)}{D},
\label{eq:app:mazur_general}
\end{equation}
where \(D\) is the Hilbert-space dimension.

For a bulk spin \(A=\sigma_j^z\), we use the three traceless conserved operators already introduced in the main text,
\begin{equation}
Q_1=b_{j-1}-\tfrac14,\qquad
Q_2=b_j-\tfrac14,\qquad
Q_3=n_{j-1}n_jn_{j+1}-\tfrac18.
\label{eq:app:Qbasis}
\end{equation}
Using
\(
\sigma_j^z=2n_j-\mathbb{I}
\)
and infinite-temperature product traces such as
\(
\mathrm{Tr}(n_i)/D=1/2
\),
\(
\mathrm{Tr}(n_in_{i+1})/D=1/4
\),
and
\(
\mathrm{Tr}(n_{j-1}n_jn_{j+1})/D=1/8
\),
one finds, for example,
\begin{equation}
\frac{\mathrm{Tr}(\sigma_j^z Q_1)}{D}=\frac18,
\qquad
\frac{\mathrm{Tr}(Q_1^2)}{D}=\frac{3}{16},
\qquad
\frac{\mathrm{Tr}(Q_1Q_2)}{D}=\frac{1}{16},
\label{eq:app:mazur_samples}
\end{equation}
with the remaining entries obtained in the same way. Substituting the resulting overlap vector and Gram matrix into Eq.~\eqref{eq:app:mazur_general} gives
\begin{equation}
M_{\sigma_j^z}=\frac35.
\label{eq:app:mazur_final}
\end{equation}
Thus at least \(60\%\) of the initial bulk polarisation is protected by the fragmented first-order dynamics.
\section{Combinatorial counting of the largest fragments}
\label{app:fibonacci}

We keep only the counting steps not written explicitly in the main text.

Let \(A_N\) denote the number of binary strings of length \(N\) with no adjacent \(1\)s. Splitting such a string by its first bit gives
\begin{equation}
A_N=A_{N-1}+A_{N-2},\qquad A_0=1,\quad A_1=2,
\label{eq:app:ANrec}
\end{equation}
because a string beginning with \(0\) is followed by any allowed length-\((N-1)\) string, while a string beginning with \(1\) must begin with \(10\) and is then followed by any allowed length-\((N-2)\) string. Therefore
\begin{equation}
A_N=F_{N+2},
\label{eq:app:ANfib}
\end{equation}
where \(F_n\) is the Fibonacci sequence with \(F_1=F_2=1\).

\subsection{Largest OBC fragment}
\label{app:fibonacci:OBC}

For the largest OBC fragment one has \(x_1=x_L=0\) together with the no-adjacent-\(1\) constraint in the bulk. Removing the fixed boundary zeros gives a bijection to an allowed open string of length \(L-2\). Hence
\begin{equation}
D_{\max}^{\mathrm{OBC}}(L)=A_{L-2}=F_L.
\label{eq:app:OBCdim}
\end{equation}

If needed, the full \(\mathbf b=\mathbf 0\) sector is obtained by summing over the four edge choices:
\begin{equation}
F_L+2F_{L-2}+F_{L-4}=F_{L+2}.
\label{eq:app:OBCsum}
\end{equation}
This identity is just repeated use of the Fibonacci recurrence.

\subsection{Largest PBC fragment}
\label{app:fibonacci:PBC}

For PBC, the no-adjacent-\(1\) strings must also satisfy the wrap-around constraint between sites \(L\) and \(1\). Splitting by the value of \(x_1\):

If \(x_1=0\), the remaining \(L-1\) sites form an allowed open string, contributing \(A_{L-1}=F_{L+1}\).

If \(x_1=1\), then both \(x_2=0\) and \(x_L=0\) are forced, so the remaining free sites form an allowed open string of length \(L-3\), contributing \(A_{L-3}=F_{L-1}\).

Therefore
\begin{equation}
D_{\max}^{\mathrm{PBC}}(L)=F_{L+1}+F_{L-1}=: \mathcal L_L,
\label{eq:app:PBCdim}
\end{equation}
where \(\mathcal L_L\) is the \(L\)-th Lucas number.

\subsection{Asymptotic scaling}
\label{app:fibonacci:strong}

Using
\begin{equation}
\mathcal L_L=\varphi^L+\psi^L,
\qquad
\varphi=\frac{1+\sqrt5}{2},
\qquad
\psi=\frac{1-\sqrt5}{2},
\label{eq:app:lucasclosed}
\end{equation}
one obtains
\begin{equation}
\frac{D_{\max}^{\mathrm{PBC}}(L)}{2^L}
\sim
\left(\frac{\varphi}{2}\right)^L,
\qquad
\frac{D_{\max}^{\mathrm{OBC}}(L)}{2^L}
\sim
\frac{1}{\sqrt5}\left(\frac{\varphi}{2}\right)^L.
\label{eq:app:strongfrag}
\end{equation}
Since \(\varphi/2<1\), the largest fragment occupies an exponentially vanishing fraction of the full Hilbert space for both boundary conditions.
\section{Third-order edge revival calculation}
\label{app:edge_pt}

We derive the third-order Magnus-cumulant Hamiltonian for an edge
spin of the open chain, establishing
Eqs.~\eqref{eq:HF2_zero}--\eqref{eq:tau_edge} in the main text.
All notation follows Sec.~\ref{sec:model}: the driven Ising chain has
Hamiltonian $H(t)=H_0(t)+H_1$ with perturbation
$H_1=-g\sum_i\sigma_i^x$, Eq.~\eqref{eq:model_H} fragmentation operator
$\mathcal{O}$, Eq.~\eqref{eq:model_O}, and triangular micromotion
function $f(t)$.
The closure $U_0(T,0)=\mathbb{I}$, Eq.~\eqref{eq:model_U0} ensures
$U_F=U^I(T,0)$, Eq.~\eqref{eq:model_Floquet}.
We work in units $\hbar=1$ except where factors of $\hbar$ clarify
dimensional analysis.

\subsection{Isolating the edge degree of freedom}
\label{app:edge:setup}

The fragmentation operator separates into a piece acting on site~1
and a remainder:
\begin{equation}
\mathcal{O}=\mathcal{O}_{\mathrm{rest}}+\sigma_1^z\,\Lambda_1,
\qquad
\Lambda_1:=h_0+J_0\sigma_2^z,
\label{eq:app:edge_Lambda}
\end{equation}
where $\mathcal{O}_{\mathrm{rest}}$ commutes with all operators on
site~1, and $\Lambda_1$ is diagonal in the computational
($\sigma^z$) basis and likewise commutes with $\sigma_1^{x,y,z}$
(since it acts only on site~2).
The interaction-picture perturbation for the left-edge spin is
therefore
\begin{equation}
V_{1,I}(t)
=-g\,e^{-if(t)\sigma_1^z\Lambda_1}\,\sigma_1^x\,
e^{+if(t)\sigma_1^z\Lambda_1},
\label{eq:app:edge_V1I}
\end{equation}
where the exponentials containing $\mathcal{O}_{\mathrm{rest}}$ cancel out.
Using the standard Pauli rotation identity
\begin{equation}
e^{-i\alpha\sigma^z}\,\sigma^x\,e^{+i\alpha\sigma^z}
=\sigma^x\cos(2\alpha)+\sigma^y\sin(2\alpha),
\label{eq:app:Pauli_rotation}
\end{equation}
we obtain
\begin{equation}
V_{1,I}(t)
=-g\bigl[\sigma_1^x\cos\theta_1(t)
+\sigma_1^y\sin\theta_1(t)\bigr],
\label{eq:app:edge_VIcompact}
\end{equation}
where the rotation angle is
\begin{equation}
\theta_1(t):=2\Lambda_1 f(t).
\label{eq:app:edge_theta}
\end{equation}
It will be convenient to introduce the unit Bloch vector
\begin{align}
&\mathbf{v}(t)
:=\bigl(-\cos\theta_1(t),\;-\sin\theta_1(t),\;0\bigr),
\label{eq:app:edge_vvec}\\
&\implies V_{1,I}(t)=g\,\mathbf{v}(t)\cdot\boldsymbol{\sigma}_1 \label{eq:V_1I}
\end{align}

\subsection{First-order edge Hamiltonian and the ``off'' condition}
\label{app:edge:first_order}

The first-order Floquet Hamiltonian is
$H_F^{(1)}=T^{-1}\!\int_0^T V_I(t)\,dt$.
For the edge contribution, the $\sigma_1^x$ and $\sigma_1^y$
coefficients require
\begin{align}
\int_0^T\!\cos\theta_1(t)\,dt
&=2\!\int_0^{T/2}\!\cos(2\Lambda_1 t)\,dt
=\frac{1}{\Lambda_1}\sin(\Lambda_1 T),
\label{eq:app:edge_cos_int}\\[4pt]
\int_0^T\!\sin\theta_1(t)\,dt
&=2\!\int_0^{T/2}\!\sin(2\Lambda_1 t)\,dt
=\frac{1}{\Lambda_1}\bigl[1-\cos(\Lambda_1 T)\bigr],
\label{eq:app:edge_sin_int}
\end{align}
where we used the tent symmetry $f(t)=f(T-t)$ to fold the
second half-period onto $[0,T/2]$.
Both integrals vanish simultaneously when
\begin{equation}
\Lambda_1 T=2\pi m,\qquad m\in\mathbb{Z}.
\label{eq:app:edge_off}
\end{equation}
We call Eq.~\eqref{eq:app:edge_off} the \textit{``off'' condition}
for the edge spin.
At $h_0=2J_0$ and $T=2\pi/J_0$, one has
$\Lambda_1 T=(2J_0+J_0\sigma_2^z)(2\pi/J_0)
=2\pi(2+\sigma_2^z)$,
which equals $2\pi$ when $\sigma_2^z=-1$ and $6\pi$ when
$\sigma_2^z=+1$; both are integer multiples of $2\pi$, so the
off condition is satisfied for every eigenvalue of $\Lambda_1$.

\subsection{Second-order cancellation by time-reversal symmetry}
\label{app:edge:second_order}

The second-order FPT is
\begin{equation}
H_F^{(2)}
=\frac{1}{2iT}\int_0^T\!dt_1\int_0^{t_1}\!dt_2\;
\bigl[V_I(t_1),\,V_I(t_2)\bigr].
\label{eq:app:edge_HF2_formula}
\end{equation}
For the single-site edge contribution, we use the Pauli commutator identity
\begin{equation}
[\mathbf{v}\cdot\boldsymbol{\sigma},\;
\mathbf{w}\cdot\boldsymbol{\sigma}]
=2i\,(\mathbf{v}\times\mathbf{w})\cdot\boldsymbol{\sigma}
\label{eq:app:Pauli_comm}
\end{equation}
and \eqref{eq:V_1I} to obtain
\begin{equation}
[V_{1,I}(t_1),\,V_{1,I}(t_2)]
=2ig^2\bigl(\mathbf{v}(t_1)\times\mathbf{v}(t_2)\bigr)
\cdot\boldsymbol{\sigma}_1.
\label{eq:app:edge_comm12}
\end{equation}
Since both $\mathbf{v}(t_1)$ and $\mathbf{v}(t_2)$ lie in the
$xy$-plane [Eq.~\eqref{eq:app:edge_vvec}], their cross product
points along~$\hat{z}$:
\begin{equation}
\mathbf{v}(t_1)\times\mathbf{v}(t_2)
=\hat{z}\,\sin\!\bigl(\theta_1(t_2)-\theta_1(t_1)\bigr),
\label{eq:app:edge_cross12}
\end{equation}
so the commutator is proportional to $\sigma_1^z$ alone.

Now we consider the following change in variables $t\mapsto T-t$ and use the fact that the micromotion function satisfies
$f(T-t)=f(t)$ for all $t\in[0,T]$, to go from the integration domain
$0<t_2<t_1<T$ to $0<T-t_1<T-t_2<T$. We now relabel the variables $(s_1,s_2):=(T-t_2,T-t_1)$ with $0<s_2<s_1<T$.
The Jacobian is $|dt_1\,dt_2|=|ds_1\,ds_2|$ (unit magnitude).
Thus the integrand transforms as
\begin{equation}
\begin{aligned}
  \sin\!\bigl(\theta_1(t_2)-\theta_1(t_1)\bigr)
\;\xrightarrow{t_i\to T-t_i}\;&
\sin\!\bigl(\theta_1(s_1)-\theta_1(s_2)\bigr)\\
&=-\sin\!\bigl(\theta_1(s_2)-\theta_1(s_1)\bigr),  
\end{aligned}
\label{eq:app:edge_TR_sign}
\end{equation}
which is the negative of the original integrand evaluated at
$(s_1,s_2)$.
Therefore the integral equals its own negative, giving
\begin{equation}
H_{F,\mathrm{edge}}^{(2)}=0.
\label{eq:app:edge_HF2_zero}
\end{equation}
This cancellation holds for \textit{any} protocol with the tent
symmetry $f(T-t)=f(t)$, not only at the off
condition~\eqref{eq:app:edge_off}.
It is equivalent to the statement that the symmetric square-wave
drive possesses a stroboscopic time-reversal symmetry that forces
the second-order cumulant to vanish for any single-site
term~\cite{Bukov2015AdvPhys}.

\subsection{Third-order FPT: general formula}
\label{app:edge:third_order}

The third-order cumulant after carrying out the FPT expansion
is~\cite{GhoshPaulSengupta2025Spin}:
\begin{equation}
\begin{aligned}
    H_F^{(3)}
&=\frac{-1}{6T}\!\int_0^T\!dt_1\!\int_0^{t_1}\!dt_2
\!\int_0^{t_2}\!dt_3\;\Bigl(
\bigl[V_I(t_1),\bigl[V_I(t_2),V_I(t_3)\bigr]\bigr]\\
&+\bigl[V_I(t_3),\bigl[V_I(t_2),V_I(t_1)\bigr]\bigr]
\Bigr).
\end{aligned}
\label{eq:app:edge_HF3_formula}
\end{equation}

We restrict to the left-edge single-site piece
$V_{1,I}(t)=g\,\mathbf{v}(t)\cdot\boldsymbol{\sigma}_1$.
The nested Pauli commutator evaluates to
\begin{equation}
\bigl[\mathbf{v}\cdot\boldsymbol{\sigma},\,
\bigl[\mathbf{w}\cdot\boldsymbol{\sigma},\,
\mathbf{u}\cdot\boldsymbol{\sigma}\bigr]\bigr]
=-4\,\bigl(\mathbf{v}\times(\mathbf{w}\times\mathbf{u})\bigr)
\cdot\boldsymbol{\sigma},
\label{eq:app:Pauli_nested}
\end{equation}
which follows from applying Eq.~\eqref{eq:app:Pauli_comm} twice.
Hence the integrand becomes
\begin{equation}
\begin{aligned}
    &\bigl[V_{1,I}(t_1),\bigl[V_{1,I}(t_2),V_{1,I}(t_3)\bigr]\bigr]\\
&+\bigl[V_{1,I}(t_3),\bigl[V_{1,I}(t_2),V_{1,I}(t_1)\bigr]\bigr]
&=-4g^3\,
\mathbf{W}_{123}\cdot\boldsymbol{\sigma}_1,
\end{aligned}
\label{eq:app:edge_integrand}
\end{equation}
where we define
\begin{equation}
\mathbf{W}_{123}
:=\mathbf{v}_1\times(\mathbf{v}_2\times\mathbf{v}_3)
+\mathbf{v}_3\times(\mathbf{v}_2\times\mathbf{v}_1),`
\mathbf{v}_j:=\mathbf{v}(t_j).
\label{eq:app:edge_W}
\end{equation}

\subsection{Projection onto Pauli components}
\label{app:edge:projection}

Note that all $\mathbf{v}_j$ lie in the $xy$-plane, the cross product
$\mathbf{v}_a\times\mathbf{v}_b=\hat{z}\sin(\theta_b-\theta_a)$
is along $\hat{z}$, and the double cross product
$\mathbf{v}_a\times(\hat{z}\sin\phi)$ lies in the $xy$-plane.
Explicitly,
\begin{equation}
\mathbf{v}_a\times(\hat{z}\sin\phi)
=\sin\phi\,\bigl(-\sin\theta_a,\;\cos\theta_a,\;0\bigr).
\label{eq:app:edge_double_cross}
\end{equation}
Applying this to both terms in $\mathbf{W}_{123}$:
\begin{align}
\mathbf{v}_1\times(\mathbf{v}_2\times\mathbf{v}_3)
&=\sin(\theta_3-\theta_2)\,
\bigl(-\sin\theta_1,\;\cos\theta_1,\;0\bigr),
\nonumber\\
\mathbf{v}_3\times(\mathbf{v}_2\times\mathbf{v}_1)
&=\sin(\theta_1-\theta_2)\,
\bigl(-\sin\theta_3,\;\cos\theta_3,\;0\bigr),
\label{eq:app:edge_Wterms}
\end{align}
where we abbreviate $\theta_j\equiv\theta_1(t_j)$.
The $\sigma_1^x$ and $\sigma_1^y$ projections of $\mathbf{W}_{123}$
are therefore
\begin{align}
W_x&:=\mathbf{W}_{123}\cdot\hat{x}
=-\sin\theta_1\sin(\theta_3-\theta_2)
-\sin\theta_3\sin(\theta_1-\theta_2),
\label{eq:app:edge_Wx}\\[4pt]
W_y&:=\mathbf{W}_{123}\cdot\hat{y}
=\cos\theta_1\sin(\theta_3-\theta_2)
+\cos\theta_3\sin(\theta_1-\theta_2).
\label{eq:app:edge_Wy}
\end{align}

\subsection{Evaluation of the $\sigma_1^y$  and $\sigma_1^x$ integral}
\label{app:edge:x_integral}

We evaluate $\int_0^T\!dt_1\!\int_0^{t_1}\!dt_2\!\int_0^{t_2}\!dt_3\;
W_y$ under the condition that $\Lambda_1T = 2m\pi$ where $m\in\mathbb Z$. Decomposing the integral into different time orderings and using the tent-like shape of $f(t)$, we obtain:
\begin{align}
    \int_0^T\!dt_1\!\int_0^{t_1}\!dt_2\!\int_0^{t_2}\!dt_3\; W_y = 0.
\end{align}

Similarly, we obtain: 
\begin{align}
    \int_0^T\!dt_1\!\int_0^{t_1}\!dt_2\!\int_0^{t_2}\!dt_3\; W_x = -\frac{3\pi T}{4\Lambda_1^2}.
\end{align}

Now we compute the third order Magnus cumulant from \eqref{eq:app:edge_HF3_formula}. We first note that \eqref{eq:app:edge_integrand} becomes \begin{equation}
\begin{aligned}
    &\bigl[V_{1,I}(t_1),\bigl[V_{1,I}(t_2),V_{1,I}(t_3)\bigr]\bigr]\\
&+\bigl[V_{1,I}(t_3),\bigl[V_{1,I}(t_2),V_{1,I}(t_1)\bigr]\bigr]
=-4g^3\,
W_x\sigma_1^x.
\end{aligned}
\end{equation}
Hence, we obtain
\begin{equation}
    \begin{aligned}
H_F^{(3)}
=\frac{-1}{6T}4g^3\,
\frac{3\pi T}{4\Lambda_1^2}\sigma_1^x = -\frac{g^3 }{2\Lambda_1^2}\sigma_1^x
\label{}
\end{aligned}
\label{eq:app:edge_HF3_final}
\end{equation}

An identical term holds at the right boundary with
$\Lambda_L:=h_0+J_0\sigma_{L-1}^z$ and $\sigma_L^x$.

\subsection{Projector decomposition at $h_0=2J_0$}
\label{app:edge:projector}

At $h_0=2J_0$:
\begin{equation}
\Lambda_1=h_0+J_0\sigma_2^z
=J_0(2+\sigma_2^z).
\label{eq:app:edge_Lambda_2J}
\end{equation}
Introduce the projectors onto the eigenvalues of $\sigma_2^z$:
\begin{equation}
\Pi_2^{\uparrow}:=\frac{1+\sigma_2^z}{2},
\qquad
\Pi_2^{\downarrow}:=\frac{1-\sigma_2^z}{2},
\qquad
\Pi_2^{\uparrow}+\Pi_2^{\downarrow}=\mathbb{I}.
\label{eq:app:edge_projectors}
\end{equation}
Then $\Lambda_1$ decomposes as
\begin{equation}
\Lambda_1=3J_0\,\Pi_2^{\uparrow}+J_0\,\Pi_2^{\downarrow},
\label{eq:app:edge_Lambda_proj}
\end{equation}
and its inverse square is
\begin{equation}
\Lambda_1^{-2}
=\frac{1}{(3J_0)^2}\,\Pi_2^{\uparrow}
+\frac{1}{(J_0)^2}\,\Pi_2^{\downarrow}
=\frac{1}{J_0^2}\Bigl(\frac{1}{9}\,\Pi_2^{\uparrow}
+\Pi_2^{\downarrow}\Bigr),
\label{eq:app:edge_Lambda_inv}
\end{equation}
where we used $\Pi_2^{\uparrow}\Pi_2^{\downarrow}=0$.
Substituting into Eq.~\eqref{eq:app:edge_HF3_final}:
\begin{equation}
\boxed{%
H_{F,\mathrm{edge}}^{(3)}
=-\frac{g^3}{2J_0^2}
\Bigl(\Pi_2^{\downarrow}
+\frac{1}{9}\,\Pi_2^{\uparrow}\Bigr)\sigma_1^x.}
\label{eq:app:edge_HF3_proj}
\end{equation}
The edge flip amplitude is $g^3/(2J_0^2)$ when the
nneighbour is down ($\sigma_2^z=-1$) and $(1/9)\times g^3/(2J_0^2)
=g^3/(18J_0^2)$ when the neighbour is up ($\sigma_2^z=+1$).
Both channels are present; neither may be neglected.

The leading boundary leakage scale is
$\|H_{F,\mathrm{edge}}^{(3)}\|\sim g^3/J_0^2$, giving the
edge memory timescale
\begin{equation}
\tau_{\mathrm{edge}}\sim\frac{J_0^2}{g^3},
\label{eq:app:edge_tau}
\end{equation}
which is \textit{independent of system size $L$}, confirming that
these edge localisation are prethermal rather than topologically
protected~\cite{ElseFendleyKempNayak2017PRX}.

\subsection{Summary of the perturbative hierarchy}
\label{app:edge:summary}

Collecting results, the edge spin at OBC experiences the following
order-by-order suppression at the fragmentation point
$h_0=2J_0$, $T=2\pi/J_0$:

\begin{center}
\begin{tabular}{lcl}
\textbf{Order} & \textbf{Result} & \textbf{Mechanism} \\
\hline
$\mathcal{O}(g)$
& $H_{F,\mathrm{edge}}^{(1)}=0$
& sinc selection rule: single-neighbor\\
& & constraint $\Rightarrow\Delta P\ne 0$ \\[3pt]
$\mathcal{O}(g^2/J_0)$
& $H_{F}^{(2)}=0$
& time-reversal symmetry of \\
& & symmetric square-wave protocol \\[3pt]
$\mathcal{O}(g^3/J_0^2)$
& $H_{F,\mathrm{edge}}^{(3)}\ne 0$
& no further protection; edge flips \\
& & via both $\Pi_2^{\downarrow}$ and
$\frac{1}{9}\Pi_2^{\uparrow}$ channels
\end{tabular}
\end{center}
\noindent
The edge dynamics is delayed by two orders relative to the bulk
PXP dynamics at $\mathcal{O}(g)$, resulting in the parametric
separation
$\tau_{\mathrm{edge}}/\tau_{\mathrm{bulk}}\sim (J_0/g)^2\gg 1$.

\section{Why the first-order Floquet term vanishes for $h_0=2nJ_0$ with $n>1$}
\label{app:freezing}

The main text states that for $h_0=2nJ_0$ with $n>1$, the first-order
Floquet Hamiltonian vanishes at $T=2\pi/J_0$. The only step worth
recording here is the classification of the allowed single-flip
$\Delta P$ values away from the resonant case $n=1$.

For a bulk spin, the neighbour sum still takes the three values
$s_{i-1}+s_{i+1}\in\{-2,0,+2\}$. Substituting $h_0=2nJ_0$ into the
local single-flip expression gives the bulk channel set
\begin{equation}
\Delta P_{\mathrm{bulk}}
\in
\bigl\{\pm 4J_0(n-1),\;\pm 4J_0 n,\;\pm 4J_0(n+1)\bigr\}.
\label{eq:app:frz_bulk_set_short}
\end{equation}
For $n>1$, none of these channels is zero. This is the essential
difference from the resonant point $n=1$, where the $n-1$ branch
produces the PXP channel.

For open chains, an edge spin has only one neighbour, so the allowed
edge channels are
\begin{equation}
\Delta P_{\mathrm{edge}}
\in
\bigl\{\pm 2J_0(2n-1),\;\pm 2J_0(2n+1)\bigr\}.
\label{eq:app:frz_edge_set_short}
\end{equation}
These are always nonzero as well. Periodic chains simply omit this
edge sector.

The freezing condition now follows immediately. At
$T=2\pi/J_0$, every bulk channel in
Eq.~\eqref{eq:app:frz_bulk_set_short} gives
$\Delta P\,T/4 = 2\pi m$ for some nonzero integer $m$, while every
edge channel in Eq.~\eqref{eq:app:frz_edge_set_short} gives
$\Delta P\,T/4 = q\pi$ for some nonzero integer $q$. Hence all
single-flip channels lie on sinc zeros, and the first-order Floquet
generator vanishes:
\begin{equation}
H_F^{(1)}=0,
\qquad
T=\frac{2\pi}{J_0},\quad h_0=2nJ_0,\quad n>1.
\label{eq:app:frz_final_short}
\end{equation}

A shorter-period drive, $T=\pi/J_0$, behaves differently for open
chains. The bulk channels are still suppressed, but the odd edge
channels give half-integer multiples of $\pi$ in the sinc argument and
therefore survive with reduced amplitude. This is why $T=\pi/J_0$
produces only partial freezing in OBC, whereas $T=2\pi/J_0$ gives
complete first-order freezing. The case $n=1$ is excluded from this
argument because the bulk channel $\Delta P=0$ reappears and generates
the constrained PXP dynamics discussed in the main text.
\begin{table}[t]
\caption{Single-flips channels at $h_0=2nJ_0$ ($n>1$, OBC).
Columns show the sinc argument $\Delta P\cdot T/4$ and whether the
sinc factor vanishes (``off'') or not (``on'').
All gaps are in units of $J_0$.}
\label{tab:app:frz_channels}
\begin{ruledtabular}
\begin{tabular}{lccccc}
& & \multicolumn{2}{c}{$T=2\pi/J_0$}
& \multicolumn{2}{c}{$T=\pi/J_0$}\\
\cline{3-4}\cline{5-6}
Channel & $|\Delta P|$
& arg & status
& arg & status \\
\hline
Bulk ($s_{i\pm 1}=\mp 1$)
& $4(n{-}1)J_0$
& $2(n{-}1)\pi$
& off
& $(n{-}1)\pi$
& off \\[3pt]
Bulk ($s_{i\pm 1}$ mixed)
& $4nJ_0$
& $2n\pi$
& off
& $n\pi$
& off \\[3pt]
Bulk ($s_{i\pm 1}=\pm 1$)
& $4(n{+}1)J_0$
& $2(n{+}1)\pi$
& off
& $(n{+}1)\pi$
& off \\[3pt]
Edge ($s_2=-1$)
& $2(2n{-}1)J_0$
& $(2n{-}1)\pi$
& off
& $\frac{(2n{-}1)\pi}{2}$
& \textbf{on} \\[3pt]
Edge ($s_2=+1$)
& $2(2n{+}1)J_0$
& $(2n{+}1)\pi$
& off
& $\frac{(2n{+}1)\pi}{2}$
& \textbf{on} \\
\end{tabular}
\end{ruledtabular}
\end{table}

\pagebreak

\bibliography{references}

\end{document}